\documentclass[epj]{svjour}
\usepackage{amsmath,amssymb,epsfig,bm,mathrsfs,slashed,color,graphicx,mathtools,isotope}%,feynmp}
\usepackage[numbers,sort&compress]{natbib}
\usepackage{filecontents}
\usepackage{tikz}
\usepackage[utf8]{inputenc}
\newcommand{\bea}{\begin{eqnarray}}
\newcommand{\eea}{\end{eqnarray}}
\newcommand{\be}{\begin{equation}}
\newcommand{\ee}{\end{equation}}

\newcommand{\vectau}{{\bm \tau}}
\newcommand{\vecrho}{{\bm \rho}}

\definecolor{red}{rgb}{0.8,0,0}
\definecolor{violet}{rgb}{0.4,0,0.4}
\definecolor{green}{rgb}{0,0.5,0.0}
\definecolor{navy}{rgb}{0.0,0.0,0.6}
\definecolor{orange}{rgb}{0.8,0.2,0.0}
\usepackage[normalem]{ulem}  % \sout{old text} for strikeout                                                 

\def\apj{ApJ~}%
\def\apjl{ApJ Lett.~}%
\def\aap{A\&A~ }%
\def\mnras{MNRAS~}%
\def\pasa{PASA}%
\def\prc{Phys.~Rev.~C~}%
\def\prd{Phys.~Rev.~D~}%
\def\sovast{Soviet~Ast.}%
\def\nphysa{Nucl.~Phys.~A}%
\def\physrep{Phys.~Rep.~}%
\usepackage[colorlinks=true]{hyperref}
\definecolor{reddish}{rgb}{0.7,0.2,0.0}
\definecolor{blueish}{rgb}{0.1,0.1,1}

\hypersetup{
  linkcolor=reddish,          % color of internal links
  citecolor=blueish,           % color of links to bibliography
urlcolor=magenta}            % color of the url           
\journalname{Eur. Phys. J. A}
\begin{document}

\title{Delta-resonances and hyperons in 
  proto-neutron stars and merger remnants}

\author{ Armen Sedrakian\inst{1,2}  and Arus Harutyunyan\inst{3,4}  
%\thanks is optional - remove next line if not needed
%\thanks{\emph{Present address:} Insert the address here if needed}%
}                     % Do not remove
\institute{
  Frankfurt Institute for Advanced Studies, Ruth-Moufang 
  str.\,\,1, D-60438 Frankfurt am Main, Germany
  \and
   Institute of Theoretical Physics, University of Wroc\l{}aw, pl. M. Borna 9, 
   50-204 Wroc\l{}aw, Poland
   \and
   Byurakan Astrophysical Observatory, Byurakan 0213, Armenia
  \and
   Department of Physics, Yerevan State University, Yerevan 0025, Armenia
}
\date{Received: 14/02/2022 / Accepted: 07/07/2022\\
\textcopyright \, 
Author(s) 2022
\\
%Communicated by N. N.
}
% The correct dates will be entered by Springer
%
\abstract{The equation of state (EoS) and composition of dense and hot
  $\Delta$-resonance admixed hypernuclear matter is studied under
  conditions that are characteristic of neutron star binary merger
  remnants and supernovas. The cold, neutrino free regime is also
  considered as a reference for the astrophysical constraints on the
  EoS of dense matter. Our formalism uses the covariant density
  functional (CDF) theory successfully adapted to include the full
  $J^P=1/2^+$ baryon octet and non-strange members of $J^P=3/2^+$
  decouplet with density-dependent couplings that have been suitably
  adjusted to the existing laboratory and astrophysical data. The effect of $\Delta$-resonances
  at finite temperatures is to soften the EoS of hypernuclear matter
  at intermediate densities and stiffen it at high densities.  At low
  temperatures, the heavy baryons $\Lambda$, $\Delta^-$,$\Xi^-$,
  $\Xi^0$ and $\Delta^0$ appear in the given order if the
  $\Delta$-meson couplings are close to those for the nucleon-meson
  couplings. As is the case for hyperons, the thresholds of
  $\Delta$-resonances move to lower densities with the increase of
  temperature indicating a significant fraction of $\Delta$'s in the
  low-density subnuclear regime. We find that the $\Delta$-resonances
  comprise a significant fraction of baryonic matter, of the order of
  $10\%$ at temperatures of the order of several tens of MeV in the
  neutrino-trapped regime and, thus, may affect the supernova and
  binary neutron star dynamics by providing, for example, a new
    source for neutrino opacity or a new channel for bulk viscosity
    via the direct Urca processes. The mass-radius relation of isentropic
    static, spherically symmetric hot compact stars is discussed.
 % end of PACS codes
} %end of abstract
\authorrunning{A. Sedrakian, A. Harutyunyan}
\titlerunning{Delta-resonances and hyperons}
\maketitle
%
%\setcounter{tocdepth}{3}
%\tableofcontents

\section{Introduction}
\label{sec:introduction}

The EoS of dense, strongly interacting matter is
the key input for an array of astrophysical simulations of compact
objects in isolation and binaries within various scenarios. A large
collection of EoS is already featured on the {\sc CopmOSE}
database~\cite{Typel:2013rza}.
Nevertheless, the need for further 
developments and export of new EoS
to this and other databases is necessary because of the strong
constraints that emerged during the recent years and will appear in
the future, notably due to the multimessenger observations of binary
neutron star (BNS) mergers, isolated nearby neutron stars in X-rays
and radiopulsars. 

The formation of hot and neutrino-rich compact objects is predicted by
numerical simulations of core-collapse supernova (SN) and BNS mergers. 
In the core-collapse supernova context, a hot proto-neutron star is formed
during the contraction of the supernova progenitor and subsequent
gravitational detachment of the remnant from the expanding
ejecta~\cite{Prakash1997,Pons_ApJ_1999,Janka_PhysRep_2007,Mezzacappa2015,Foglizzo2015,Connor2018ApJ,Burrows2020MNRAS,Pascal2022}.
A transient formation of dense hot matter arises in the case when
the progenitor mass is large (typically tens of solar masses)  and the
matter collapses into a black
hole~\cite{Sumiyoshi2007,Fischer2009,OConnor_2011,Schneider2020}.

Numerical simulations indicate that BNS mergers produce hot and dense
interacting matter in the post-merger
phase~\cite{Shibata_11,Rosswog2015,Baiotti2017,Baiotti:2019sew,Hanau2019}.
The outcome of a merger depends on the combined masses of merged
objects and may result both in a black hole and a stable neutron star.
In any case, a transient hot object is formed and, therefore, the 
spectrum of gravitational waves emitted in this phase (which can be
observed with advanced gravitational wave instruments) will carry
imprints of the EoS of hot and dense matter. This EoS also determines
the stability of the remnant object and thus the outcome of the
transient evolution~\cite{Khadkikar:2021yrj} as well as the efficacy
of dissipative
processes~\cite{Alford2019a,Alford2019b,Alford2020a,Alford2021a,Alford2021b,Alford2021c,Most2022MNRAS,Celora:2022nbp} that should be included~\cite{Most2022MNRAS} in the frequently employed
ideal hydrodynamics simulations.
%,

The local properties of matter in the “hot” stage of evolution in the
above astrophysical contexts is characterized by the density,
temperature (or entropy), and the lepton fractions for electrons and
$\mu$-ons. The EoS in this stage depends on multiple parameters which
can be compared to the simpler one-parametric EoS of cold and
$\beta$-equilibrated matter. Since many nuclear and astrophysical
constraints are placed on the cold EoS, it is mandatory to study this
limit in parallel with the finite-temperature EoS. The cold and hot 
regimes also differ in the fact that in the hot case neutrinos are 
trapped above the temperature $T_{\rm tr}\simeq 5$~MeV; in this regime,
the neutrino mean-free path is shorter than the size of the
star~\cite{Alford2018b}. As we will discuss below the composition
of matter is strongly affected by the prescription for the neutrino
fractions (via the fixation of the lepton number). 

In this work, we report an extension of our previous study of hot
hypernuclear matter~\cite{Sedrakian2021Univ} which builds upon the
work of Ref.~\cite{Colucci2013} to include non-strange
$J^P=\frac{3}{2}^+$ members of the baryons decuplet -- the
$\Delta$-resonances.  Our numerical implementation is based on the
code of Ref.~\cite{Colucci2013} supplemented with hidden strangeness
$\sigma^*$ and $\phi$ mesons which account for the interactions
amongst hyperons~\cite{Sedrakian2021Univ}. In the hypernuclear sector,
we will adopt the parameter set already discussed in
Refs.~\cite{Lijj2018b,Lijj2019,Li2020PhRvD} in the case of
zero-temperature EoS. In the nucleonic sector, we will use CDF
parameters corresponding to the DDME2
parameterization~\cite{Lalazissis2005}. Heavy baryons have been
studied in the zero-temperature limit in recent years because of the
emerging new astrophysical and laboratory
constraints~\cite{Schurhoff2010,Drago_PRC_2014,Cai_PRC_2015,Zhu_PRC_2016,Kolomeitsev2017,Sahoo_PRC_2018,Lijj2018b,Lijj2019,Ribes_2019},
for a review see~\cite{Sedrakian2021}.  The study of the hypernuclear
matter with $\Delta$-resonance admixture at finite tem\-pe\-ra\-tures
begun recently~\cite{Malfatti:2019tpg,Spinella2020:WSBook,Raduta:2020fdn}.
These references employed CDFs with density-dependent couplings, as we
do below.  However, our study differs from these works by (a) the
parametrization of the CDF in the nuclear and/or hypernuclear sectors;
(b) the mesonic content of the Lagrangian.  Specifically
Refs.~\cite{Malfatti:2019tpg,Spinella2020:WSBook,Raduta:2020fdn}
include $\sigma, \omega,$ and $\rho$ mesons only while we will include
in addition the hidden strangeness $\sigma^*$ and $\phi$ mesons.
  
  This work is organized as follows. In Section~\ref{sec:EoS_npe+H+D}
  we discuss the main elements of CDF approach at finite temperatures. 
  The general purpose EoS is then specified for the scenarios of SN 
  and BNS mergers in Sec.~\ref{sec:thermodynamics}. We present the 
  numerical results on the EoS and composition in 
  Section~\ref{sec:composition_EoS}. The mass-radius (hereafter 
  $M$-$R$) relation of static cold and isentropic, hot compact stars 
  and the astrophysical constraints are discussed in Section~\ref{sec:MR}. 
  In Section~\ref{sec:conclusions} we provide a summary of our main
  findings. We use the natural (Gaussian) units with $\hbar= c=k_B=1$,
  and the metric signature $g^{\mu\nu}={\rm diag}(1,-1,-1,-1)$.

\section{Relativistic density functional with  density-dependent couplings}
\label{sec:EoS_npe+H+D}

\subsection{Equation of state}
\label{sec:Hyper_DDME2}

The Lagrangian of the stellar matter is given by 
% --------------------------------------------------------
\begin{equation}
  \label{eq:Lagrangian}
  \mathscr{L} = \mathscr{L}_b + \mathscr{L}_d + \mathscr{L}_m +  \mathscr{L}_\lambda
  +  \mathscr{L}_{\rm em},
\end{equation}
% --------------------------------------------------------
where the $J_B^P = \frac{1}{2}^+$  baryon Lagrangian is given by 
% --------------------------------------------------------
\begin{eqnarray}
  \label{eq:Lagrangian_b} 
  \mathscr{L}_b \, &=&\,  \sum_b\bar\psi_b\bigg[\gamma^\mu 
  \left(i\partial_\mu-g_{\omega b}\omega_\mu
  -g_{\phi b}\phi_\mu - \frac{1}{2} g_{\rho
  b}\vectau\cdot\vecrho_\mu\right) \nonumber\\
  &-& (m_b - g_{\sigma b}\sigma 
  - g_{\sigma^* b}\sigma^*)\bigg]\psi_b ,
\end{eqnarray}
% --------------------------------------------------------
where the $b$-sum is over the $J_B^P = \frac{1}{2}^+$ baryon octet
$b\in (n,p,\Lambda, \Xi^{0,-},\Sigma^{0,\pm})$, $\psi_b$ are the Dirac
fields of the octet with masses $m_b$.  The mesonic fields included in
theory are $\sigma,\sigma^*,\omega_\mu,\phi_\mu$, and $\vecrho_\mu$
with meson-baryon couplings $g_{mb}$ where $m$ index runs over the
mesons $m\in (\sigma, \omega, \rho, \sigma^*, \phi)$. Later on we will
specify the theory to the case where the couplings are
density-dependent.  The strange mesons $\sigma^*$ and $\phi$ couple
only to hyperons.  The second term in Eq.~\eqref{eq:Lagrangian} stands
for the contribution of the non-strange $J=\frac{3}{2}^+$ members of
the baryons decuplet which is the quartet of $\Delta$-resonances
$d \in (\Delta^-,\Delta^0,\Delta^+,\Delta^{++})$ and is given
explicitly by
% --------------------------------------------------------
\begin{eqnarray}
  \mathscr{L}_d = \sum_{d}\bar\psi^{\nu}_{d}
  \bigg[\gamma^\mu \left(i\partial_\mu-g_{\omega d}\omega_\mu
  - \frac{1}{2} g_{\rho d}\vectau\cdot\vecrho_\mu\right)\nonumber\\
                   - (m_d- g_{\sigma d}\sigma)\bigg]\psi_{d\nu} ,
\end{eqnarray}
% --------------------------------------------------------
where the $d$-summation is over the resonances described by the Rarita-Schwinger fields $\psi_{d\nu}$.

The mesonic Lagrangian is given by
% --------------------------------------------------------
\begin{eqnarray}
\mathscr{L}_m &=& \frac{1}{2}
\partial^\mu\sigma\partial_\mu\sigma-\frac{m_\sigma^2}{2} \sigma^2 -
                  \frac{1}{4}\omega^{\mu\nu}\omega_{\mu\nu}
                  + \frac{m_\omega^2}{2}
                  \omega^\mu\omega_\mu \nonumber\\
  &-& \frac{1}{4}\vecrho^{\mu\nu}\cdot \vecrho_{\mu\nu}
               + \frac{m_\rho^2}{2} \vecrho^\mu\cdot\vecrho_\mu 
               +\frac{1}{2}
\partial^\mu\sigma^*\partial_\mu\sigma^*-\frac{m_\sigma^{*2}}{2}
  \sigma^{*2}\nonumber\\
  &-&
\frac{1}{4}\phi^{\mu\nu}\omega_{\mu\nu} + \frac{m_\phi^2}{2}
               \phi^\mu\phi_\mu,
\end{eqnarray}
% --------------------------------------------------------
where $m_{\sigma}$, $m_{\sigma^*}$, $m_{\omega}$, $m_{\phi}$ and 
$m_{\rho}$ are the meson masses. 
The field-strength tensors for vector
fields are given by 
% --------------------------------------------------------
\begin{eqnarray}             
\omega_{\mu \nu}  &=& \partial_{\mu}\omega_{\nu} - \partial_{\mu}\omega_{\nu} ,\\
\phi_{\mu \nu} & =& \partial_{\mu}\phi_{\nu} - \partial_{\mu}\phi_{\nu} ,\\
\boldsymbol{\rho}_{\mu \nu}  &=& \partial_{\nu}
\boldsymbol{\rho}_{\mu} - \partial_{\mu}\boldsymbol{\rho}_{\nu}.
\end{eqnarray}
% --------------------------------------------------------
The leptons will be assumed non-interacting and are described by
the free-field Dirac Lagrangian 
% --------------------------------------------------------
\begin{eqnarray}
\label{eq:Lagrangian_leptons}
\mathscr{L}_\lambda\,=\, \sum_{\lambda}\bar\psi_\lambda(i\gamma^\mu\partial_\mu -
      m_\lambda)\psi_\lambda,
\end{eqnarray}
% --------------------------------------------------------
where $\psi_\lambda$ are leptonic fields and $m_\lambda$ are their
masses. In the case of cold stellar matter the lepton index
$\lambda \in (e,\mu)$ runs over electrons and $\mu$-ons and their
antiparticles, whereas $\tau$-leptons can be neglected because of
their large mass.  For temperatures above trapping temperature
$T_{\rm tr}=5$~MeV electron and $\mu$-on neutrinos are trapped (the
details of trapping depend on the local density of matter and its
composition). In that case, the three flavors of left-handed neutrinos
(assuming Standard Model particle content) need to be included in the
Lagrangian~\eqref{eq:Lagrangian_leptons}.  In this work, we neglect
electromagnetism and drop the term $\mathscr{L}_{\rm em}$ from the
Lagrangian~\eqref{eq:Lagrangian}. It has been included 
elsewhere to accommodate the possibility of extremely large
magnetic fields in compact stars,
see~\cite{Sinha2013,Thapa:2020ohp,Dexheimer:2021sxs}
and references therein.

Having defined the Lagrangian \eqref{eq:Lagrangian} of the system we
proceed to evaluate the partition function of the system. The
evaluation is simplified by the fact that we consider a stationary
system in the infinite limit, i.e., the time and space variations of
the fields can be neglected. The partition function is evaluated in
the mean-field approximation by keeping only the Hartree terms. With
these approximations, the pressure and energy density are given by
%------------------------------------------------------------------------
\begin{eqnarray}
  \label{eq:P}
  P  &=&  P_m+P_b+P_d+P_{\lambda} + P_{r},\\
    \label{eq:E}
          {\cal E} &=&  {\cal E}_m+{\cal E}_d+{\cal E}_d+{\cal E}_{\lambda},
\end{eqnarray}
where the contributions due to mesons and  $J_B^P =
\frac{1}{2}^+$-baryons are given by
\begin{eqnarray}
    \label{eq:P_m}
P_m &=& - \frac{m_\sigma^2}{2} \sigma^2 -\frac{m_\sigma^{*2}}{2} \sigma^{*2}
          + \frac{m_\omega^2}{2} \omega_0^2 +  \frac{m_\phi^2 }{2} \phi_0^2
        + \frac{m_\rho^2 }{2} \rho_{03}^2,\nonumber\\ \\
% ------------------------------------------------------------------------
    \label{eq:E_m}
{\cal E}_m &=& \frac{m_\sigma^2}{2} \sigma^2 +\frac{m_\sigma^{*2}}{2} \sigma^{*2} +
             \frac{m_\omega^2 }{2} \omega_0^2 
                          +  \frac{m_\phi^2 }{2} \phi_0^2
                          + \frac{m_\rho^2 }{2} \rho_{03}^2,\\
% ------------------------------------------------------------------------
    \label{eq:P_b}
  P_b&=& \sum_{b} \frac{g_{b}}{6\pi^2}\int_0^{\infty}\!
  \frac{dk\, k^4 }{E_k^{b}}
     \left[f(E_k^{b}-\mu_{b}^*)+f(E_k^{b}+\mu_{b}^*)\right],\nonumber\\\\
% ------------------------------------------------------------------------  
    \label{eq:E_b}
    {\cal E}_b &=& \sum_{b} \frac{g_{b}}{2\pi^2}  
\int_0^{\infty}\! dk \, k^2 E_k^{b} \left[f(E_k^{b}-\mu_{b}^*)
                   +f(E_k^{b}+\mu_{b}^*)\right] ,\nonumber\\
\end{eqnarray}
% ------------------------------------------------------------------------
where $f(E) = [1+\exp(E/T)]^{-1}$ is the Fermi distribution function,
$g_b=2J_b+1 = 2 $ is the spin ($J_b = 1/2$) degeneracy factor of the
baryon octet. The expressions for $P_d$ and ${\cal E}_d$ follow
from  \eqref{eq:P_b} and  \eqref{eq:E_b} via a replacement of the
indices $b$ by $d$ and taking into account that the spin degeneracy
factor is $g_d = 4 $ for $\Delta$-resonances.
The lepton contribution is given by
%------------------------------------------------------------------------
\begin{eqnarray}
  P_{\lambda} &=&
   \sum_{\lambda}
  \frac{g_{\lambda}}{6\pi^2}\int_0^{\infty}\!
   \frac{dk\, k^4 }{E_k^\lambda}
   \left[f(E_k^{\lambda}-\mu_\lambda)+f(E_k^{\lambda}+\mu_\lambda)\right],\nonumber\\
  \\
  {\cal E}_{\lambda} &=&  \sum_{\lambda}                         \frac{g_{\lambda}}{2\pi^2}  \int_0^{\infty}\!\!
  dk\, k^2E_k^\lambda\left[f(E_k^\lambda-\mu_\lambda)
+f(E_k^\lambda+\mu_\lambda)\right], \nonumber\\
\end{eqnarray}
%------------------------------------------------------------------------
where the degeneracy factor $g_\lambda= 2J_{\lambda}+1 $ is equal 2
for electrons and $\mu$-ons and $1$ for neutrinos of all flavors.  The
single-particle energies of baryons and $\Delta$-resonances (which
include interactions) are given by $E_k^{b} = \sqrt{k^2+m^{*2}_{b}}$
and $E_k^{d} = \sqrt{k^2+m^{*2}_{d}}$, respectively, where the
corresponding effective (Dirac) masses are given by
%------------------------------------------------------------------------
\begin{equation}
  m_{b}^* = m_b - g_{\sigma b}\sigma - g_{\sigma^*b}\sigma^*,\quad
  m_{d}^* = m_d - g_{\sigma d}\sigma,
\end{equation}
% ------------------------------------------------------------------------
where the mesonic fields now correspond to their mean-field values, see
Eq.~\eqref{eq:m_sigma}-\eqref{eq:m_rho} below.  Leptons are treated as
non-interacting gas and their kinetic energies are given by
$E_k^{\lambda}=\sqrt{k^2+m_\lambda^2}$, where $m_\lambda$ is given by
the free mass of electron or $\mu$-on and is assumed vanishingly small in
the case of neutrinos.

For contact interactions, the mesonic mean-fields shift the value of the baryon
and $\Delta$-resonance non-interacting chemical potentials $\mu_{b}$
and $\mu_{d}$ to
%------------------------------------------------------------------------
\begin{eqnarray}
\label{eq:mu_eff}
&& \mu_{b}^* = \mu_{b}- g_{\omega b}\omega_{0} - g_{\phi b}\phi_{0} 
   - g_{\rho b}  \rho_{03} I_{3b} - \Sigma^{r}, \\
\label{eq:mu_eff_d}
  && \mu_{d}^* = \mu_{d}- g_{\omega d}\omega_{0} 
- g_{\rho d}  \rho_{03} I_{3d} - \Sigma^{r}, 
\end{eqnarray}
%------------------------------------------------------------------------
where  $I_{3b/3d}$ is the third component 
of isospin of baryons/$\Delta$-resonances and the 
rearrangement self-energy $\Sigma^r$ is given by
%------------------------------------------------------------------------
\begin{eqnarray}
  \Sigma^r&=&\sum_{b,d} \left(
  \frac{\partial g_{\omega b}}{\partial n_b} \omega_0 n_b+
 \frac{\partial g_{\rho b}}{\partial n_b}  I_{3b} \rho_{03} n_b+
              \frac{\partial g_{\phi b}}{\partial n_b}  \phi_0 n_b
              \right.\nonumber\\
  &-&\left.
  \frac{ \partial g_{\sigma b}}{\partial n_b} \sigma n_b^s
  -  \frac{ \partial g_{\sigma^* b}}{\partial n_b} \sigma^* n_b^s
  + b\leftrightarrow d\right).
\end{eqnarray}
% ------------------------------------------------------------------------
This quantity adds a contribution to the pressure in a manner that
guarantees the thermodynamical consistency (specifically the energy
conservation and fulfillment of the Hugenholtz--van Hove theorem).
The true pressure is given 
by Eq.~\eqref{eq:P} where the rearrangement term is
%------------------------------------------------------------
\bea\label{eq:thermodynamic_p} P_r = n_B\Sigma_{r},
\eea
% ------------------------------------------------------------
where $n_B$ is the net baryon density.  It can be verified that the
contribution from the rearrangement self-energy guarantees the
validity of the thermodynamic relation
%------------------------------------------------------------
\be
P = n_B^2 \frac{\partial}{\partial n_B}\left(\frac{{\cal E}}{n_B}\right).
\ee
%------------------------------------------------------------

The expectation values of mesons in the mean-field and infinite system
approximations are given by
%------------------------------------------------------------------------
\begin{eqnarray}
  \label{eq:m_sigma}
      &&  m_{\sigma}^2\sigma = \sum_{b} g_{\sigma b}n_{b}^s + \sum_{d}
         g_{\sigma d}n_{d}^s, \\
    \label{eq:m_sigmas}
 && m_{\sigma^{*}}^2 \sigma^{*} = \sum_{b} g_{\sigma^{*}   b}
    n_{b}^s,\\
      \label{eq:m_omega}
  && m_{\omega}^2\omega_{0}= \sum_{b} g_{\omega b}n_{b} +\sum_{d}
     g_{\omega d}n_{d}, \\
        \label{eq:m_phi}
      && m_{\phi}^2\phi_{0}= \sum_{b} g_{\phi b}n_{b} ,\\
        \label{eq:m_rho}
&&  m_{\rho}^2\rho_{03}= \sum_{b} g_{\rho b}
  n_{b}  I_{3b}+ \sum_{d} g_{\rho d} n_{d}I_{3d},
\end{eqnarray}
%------------------------------------------------------------------------
where the scalar and baryon (vector) number densities are defined for
the baryon octet and  $\Delta$-resonances as
%------------------------------------------------------------------------
\bea n_{b}^s&=& \langle\bar{\psi}_b \psi_b\rangle,
\quad n_{b}= \langle\bar{\psi}_b \gamma^0 \psi_b\rangle,\\
n_{d}^s&=& \langle\bar{\psi}_{d\nu} \psi^\nu_d\rangle,\quad
n_{d}= \langle\bar{\psi}_{d\nu} \gamma^0 \psi^\nu_d\rangle,
\eea
%------------------------------------------------------------------------
respectively.
%------------------------------------------------------------------------
The explicit expressions of these expectation values at finite
temperatures can be computed in a standard way, and are given in the
case of spin-$J\frac{1}{2}^+$ by
%-------------------------------------------------------
\bea\label{eq:density_b}
n_{b} &=& \frac{g_b}{2\pi^2}\!\!\int_0^\infty\!   k^2dk
\left[f(E^b_k-\mu_b^*)-f(E^b_k+\mu_b^*)\right],\\
\label{eq:density_s}
n_{b}^s &=& \frac{g_b}{2\pi^2}\!\!\int_0^\infty\! \frac{k^2dk\, m^*_b}{E_k^b} 
\left[f(E^b_k-\mu_b^*)+f(E^b_k+\mu_b^*)\right].\quad
\eea 
% -------------------------------------------------------
The expressions for the $\Delta$-resonances are obtained upon
exchange $b\leftrightarrow d$.

\subsection{Fixing couplings}
\label{sec:couplings}

% ---------------------
\begin{table*}[t]
  \centering
\caption{The values of parameters of the DDME2 CDF.}
{
\begin{tabular}{ccccccc}
\hline \hline
Meson ($i$) & $m_i$ (MeV) & $a_{i}$ & $b_{i}$ & $c_{i}$ & $d_{i}$ & $g_{iN}$ \\
\hline
$\sigma$ & 550.1238 & 1.3881 & 1.0943 & 1.7057 & 0.4421 & 10.5396 \\
$\omega$ & 783 & 1.3892 & 0.9240 & 1.4620 & 0.4775 & 13.0189 \\
$\rho$ & 763 & 0.5647 & -- & -- & -- & 7.3672 \\
  \hline
\end{tabular}
}
\label{tab:1}
\end{table*}
% -------

In this work,  we continue to employ a model with density-dependent
couplings that depend on the net baryon density $n_B$. The
influence of the temperature on the effective interactions in the
theory (as well as possible contributions from fluctuations) is thus
neglected. The density-dependence of the nucleon-meson couplings is
%------------------------------------------------------------
\begin{equation}
 g_{iN}(n_B) = g_{iN}(n_{\rm sat})h_i(x),
\end{equation}
% -----------------------------------------------------------
where $n_{\rm sat}=0.152$~fm$^{-3}$ is the saturation density,
$x = n_B/n_{\rm sat}$ and
% ----------------------------------------------------------
\begin{eqnarray} \label{eq:h_functions}
  h_i(x) &=&\frac{a_i+b_i(x+d_i)^2}{a_i+c_i(x+d_i)^2},~i=\sigma,\omega,\\
  \label{eq:h_function_rho}
h_\rho(x) &=& e^{-a_\rho(x-1)}.
\end{eqnarray}
% ---------------------
The five constraints $h_{i }(1)=1$, $h_i^{\prime\prime}(0)=0$ and
$h^{\prime\prime}_{\sigma}(1)=h^{\prime\prime}_{\omega }(1)$ allow
one to reduce the number of free parameters in isoscalar-scalar and
iso-scalar-vector sector to three.  In the nucleonic (hereafter $N$)
sector, the parameters of the model are fixed from the nuclear
phenomenology and properties of selected nuclei. We adopt the DDME2
model \cite{Lalazissis2005} with the couplings and other parameters
defined in Table~\ref{tab:1}.  For a review of the theory that uses
density-dependent couplings for the meson-baryon interactions see, for
example, \cite{Typelparticles2018}.

Let us turn to the hyperonic (hereafter $Y$) sector. We follow the
established procedure to fix the couplings of the vector mesons
according to the SU(6) spin-flavor symmetric model~\cite{Swart1963}
and adjust the scalar meson couplings to reproduce the values of the
phenomenological potential depths of various hyperons at the
saturation density in isospin symmetrical nuclear matter.
% ----------------------------
\begin{table}[t]
\centering
\caption{The ratios of the couplings of hyperons and
  $\Delta$-resonances to mesons to those of nucleons in our model.}
%---------------------------
\begin{tabular}{cccccc}
    \hline
 $b\backslash R$  & $R_{\omega b}$  & $R_{\phi b}$ & $R_{\rho b}$
 & $R_{\sigma b}$  &$R_{\sigma^* b}$\\
  \hline
$\Lambda$ &   2/3 &  $-\sqrt{2}/3$    & 0 & 0.6106 & 0.4777 \\
$\Sigma $ &   2/3 &  $-\sqrt{2}/3$    & 2 & 0.4426 & 0.4777 \\
  $\Xi$     &   1/3 &  $-2\sqrt{2}/3$   & 1 & 0.3024 & 0.9554\\
  $\Delta^{-}$     &   1 &  0  & 1 & 1 & 0\\
  $\Delta^{0}$     &   1 &  0   & 1 & 1 & 0\\
  $\Delta^{+}$     &   1 &  0   & 1 & 1 & 0\\
  $\Delta^{++}$     &  1 &  0   & 1 & 1 & 0\\
\hline
\end{tabular}
\label{tab:2}
\end{table}
%---------------------------
Quantitatively one defines the ratios of hyperonic couplings to the
corresponding nucleonic couplings, i.e.,
$R_{i Y}=g_{i Y}/g_{i N} $ for $i=\{\sigma,\omega,\rho\}$ and
$R_{\sigma^* Y} =g_{\sigma^* Y}/g_{\sigma N} $,
$R_{\phi Y}=g_{\phi Y}/g_{\omega N} $. The values of corresponding
ratios are listed in Table~\ref{tab:2}. The ratio 
$R_{\sigma\Lambda}$ for $\Lambda$-hyperons~\cite{Lijj2018b} is numerically 
close to the value determined from fits to $\Lambda$-hypernuclei data~\cite{Dalen2014}.
The commonly considered range of potentials for $\Sigma$ 
and $\Xi$ hyperons, in the sense defined above, is given by 
% ------------------------------
\bea
&&  -10\le U_{\Sigma}(n_{\rm sat}) \le 30 ~{\rm MeV},\\
&& -24\le U_{\Xi}(n_{\rm sat}) \le 0~{\rm MeV}. 
\eea
% ------------------------------
The lower value of the range $U_{\Xi}(n_{\rm sat}) $ was obtained from
the analysis of the $\Xi+p \to \Lambda\Lambda$ two-body capture events
in $^{12}$C and $^{14}$N emulsion nuclei~\cite{Friedman2021PhLB}; more
shallow results were obtained from the analysis of the
$^{9}$Be($K^{-},K^{+}$) reaction, specifically, $U_{\Xi}(n_{\rm sat})
=-17$~MeV~\cite{Harada2021PhRvC}, and on the bases of the (2+1)-flavor
lattice QCD simulations close to the physical point by the Lattice19
collaboration~\cite{Sasaki:2019qnh}. Our adopted values of the
couplings match those used previously by Ref.~\cite{Lijj2018b}. The
remaining parameters in the hyperonic sector, which determine the
density-dependence of the couplings, are the same as in the nucleonic
sector. In particular, the hidden strangeness mesons are assigned
masses $m_{\sigma^*} = 980$~MeV and $m_{\phi} = 1019.45$~MeV, and the
density-dependence of their couplings coincides with the ones of
$\sigma$- and $\omega$-mesons, respectively.

Finally, let us turn to the $\Delta$-resonance (hereafter $\Delta$) matter.
The information on the $\Delta$-potential in the isospin symmetric
nuclear matter is available from the analysis of the scattering of
electrons and pions off nuclei and from the simulations of the heavy
ion collisions.  The isovector meson-$\Delta$-resonance couplings are
less explored.  Recent work suggests for the ratios
the ranges~\cite{Lijj2018b,Li2019PhysRevC}
%---------------------------------------------------------------
\begin{equation}
R_{\rho\Delta} = 1,  \quad  0.8 \le R_{\omega\Delta} \le 1.6,     \quad            
R_{\sigma\Delta} = R_{\omega\Delta} \pm 0.2.
\end{equation}
% ---------------------------------------------------------------
In the following we adopt representative values
% ---------------------------------------------------------------
\bea\label{eq:R}
R_{\rho\Delta} = R_{\omega\Delta} = R_{\sigma\Delta} =1.
\eea
% ---------------------------------------------------------------
Note that within a certain range of the parameters the
$\Delta$-admixed matter undergoes spinodal
instability~\cite{Raduta2021PhLB}; for fixed
$R_{\rho\Delta} = R_{\sigma\Delta} =1$ this occurs for
$R_{\omega\Delta} \le 0.8$, therefore the choice \eqref{eq:R} avoids
such instabilities.

\subsection{Adapting CDF to conditions in supernovas and merger remnants}
\label{sec:thermodynamics}

Consider next matter composed of baryon octet, $\Delta$-reso\-nances,
and leptons. If equilibrium is established in matter with respect to
the weak processes, then the following relations for the chemical
potentials of the species hold
% -----------------------------------------------------------------
\begin{eqnarray}
\label{eq:c1}
  &&\mu_{\Lambda}=\mu_{\Sigma^0}=\mu_{\Xi^0}=\mu_{\Delta^0}=\mu_n=\mu_B,\\
\label{eq:c2}
 &&  \mu_{\Sigma^-}=\mu_{\Xi^-}=\mu_{\Delta^-}=\mu_B-\mu_Q,\\
\label{eq:c3}
  &&\mu_{\Sigma^+}=\mu_{\Delta^+}=\mu_B+\mu_Q,\\
\label{eq:c4}
 && \mu_{\Delta^{++}}=\mu_B+2\mu_Q,
\end{eqnarray}
% -----------------------------------------------------------------
where $\mu_B$ and $\mu_Q=\mu_p-\mu_n$ are the baryon and charge
chemical potentials.  The net charge of baryons is given by the sum 
% -----------------------------------------------------------------
    \begin{eqnarray}
    &&  n_p+n_{\Sigma^+}+2n_{\Delta^{++}}\nonumber\\
&&\hspace{1cm}+n_{\Delta^{+}}-(n_{\Sigma^-}+n_{\Xi^-}+n_{\Delta^-})=n_Q.
\end{eqnarray}
% -----------------------------------------------------------------
Next we define the dimensionless baryon and lepton charge densities
via $Y_Q = n_Q/n_B$, $Y_{e,\mu} = (n_{e,\mu}-n_{e^+,\mu^+})/n_B$,
where $e^+$ refers to the positron and $\mu^+$ -- to the
anti-$\mu$-on. Then, the charge neutrality condition in terms of these new
quantities can be written
% -----------------------------------------------------------------
\begin{equation}
Y_Q = Y_e + Y_{\mu}.
\end{equation}
% -----------------------------------------------------------------
When neutrinos are trapped in matter, i.e., are in thermal equilibrium
characterized by a distribution function at matter temperature, 
 the quantities that are fixed are the lepton numbers
$Y_{L,e} = Y_{e} + Y_{\nu_e}$ and $Y_{L,\mu} = Y_{\mu} + Y_{\nu_\mu}$
of the electron and $\mu$-on families, respectively, which are conserved
separately and, therefore, are associated with lepton-number chemical
potentials $\mu_{L,e} $ and $\mu_{L,\mu}$.  In the free-streaming
(untrapped) neutrino regime the neutrino chemical potentials vanish
and the lepton chemical potentials are equal to the charge chemical
potential up to the sign. Thus, we have
% -----------------------------------------------------------------
\begin{eqnarray}
\label{eq:mu_condition}
  \mu_e &=& \mu_\mu =-\mu_Q = \mu_n-\mu_p, \quad \textrm{(free streaming)}\\
  \mu_e& = & \mu_{L,e} - \mu_Q, \quad \mu_{\mu} =  \mu_{L,\mu} -
  \mu_Q.   \quad \textrm{(trapped)}
\end{eqnarray}
% --------------------------- --------------------------------------
The thermodynamical conditions depend on the astrophysical scenario
under consideration. As well known, neutrinos are trapped when their
mean-free path is shorter than the size of the system (roughly the
stellar radius) ~\cite{Alford2019a,Alford2019b,Alford2020a}. The
electron and $\mu$-on neutrino spheres (with their surfaces defined by
the location of the last neutrino scattering) are not identical and,
therefore, the trapping regimes may depend on the lepton family.

We will fix below the lepton number in each family separately,
assuming that the neutrino oscillations are neglected and will neglect
$\tau$-leptons as they are too massive to be relevant.

In the case of BNS mergers, the initial conditions
correspond to two cold neutron stars, which are dominated by the
neutron component. In this case, our working assumption is
% ----------------------------------------------------------------
\bea\label{eq:Y_BNS}
Y_{L,e} = Y_{L,\mu}= 0.1,
\eea
% ----------------------------------------------------------------
which is consistent with the lepton abundances in the pre-merger
neutron stars.

For supernova matter the predicted electron and $\mu$-on lepton numbers
are typically~\cite{Prakash1997,Malfatti:2019tpg}
%----------------------------------------------------------------
\bea\label{eq:Y_SN}
Y_{L,e} = 0.4, \quad Y_{L,\mu} = 0.
\eea
% ----------------------------------------------------------------
However, note that the electron fraction may vary significantly along
the supernova profile in a time-dependent manner. Furthermore,
$\mu$-onization in the matter can lead to a small fraction of
$\mu$-ons (of the order $10^{-3}$)~\cite{Bollig:2017lki,Guo:2020tgx}
which we will neglect here.

We will show below isentropic results for the entropy per baryon $S/A
= 1$, which is a representative value of the entropy of the core of a
BNS remnant~\cite{Perego2019EPJA} and of a core region of a supernova
and proto-neutron star~\cite{Nakazato2022ApJ}.  In both cases the
dense core of the star maintains its low entropy, whereas the outer
layers are heated by shock(s) and dissipation and may reach a larger
value of $S/A$. The sensitivity of the EoS and composition on the
value $S/A$ has been quantified in Ref.~\cite{Raduta:2020fdn} who used
a somewhat different density functional.

\section{Composition and EoS of hot $NY\Delta$ matter}
\label{sec:composition_EoS}

We have extended the numerical code for the computation of finite
temperature hypernuclear matter, as presented in
Ref.~\cite{Colucci2013}, to include $\Delta$-resonances. The numerical
procedure is based on a self-consistent solution of the equations for
the meson fields \eqref{eq:m_sigma}-\eqref{eq:m_rho} and the scalar
and baryon densities \eqref{eq:density_b} and \eqref{eq:density_s} for
fixed values of temperature and density or, alternatively,
  entropy per baryon $S/A$ and density. In the neutrino-trapped
regime the lepton numbers $Y_{L,e}$ and $Y_{L,\mu}$ are fixed to
values that are characteristic either for supernova or BNS merger
physics, see Sec.~\ref{sec:thermodynamics}. 

\begin{figure}[t] 
\begin{center}
\includegraphics[width=1.1\linewidth,keepaspectratio]{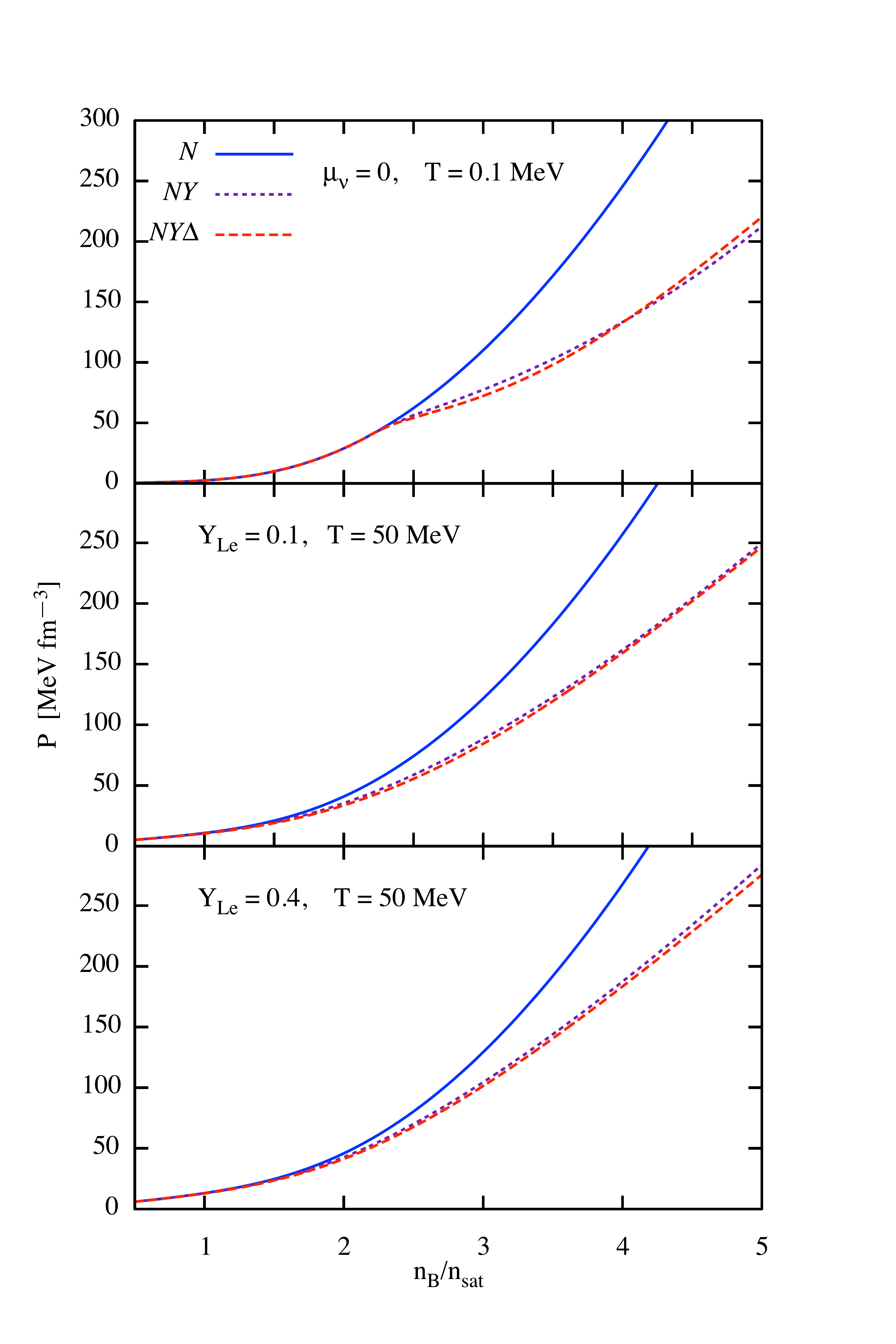}
\caption{Pressure as a function of the baryon density normalized by
  the saturation density $n_{\rm sat}$.  The upper panel labeled
  $\mu_\nu=0$ assumes neutrino-free $\beta$-equilibrium matter in the
  cases of purely nucleonic (labeled $N$), hyperonic (labeled $NY$)
  and $\Delta$-admixed hyperonic (labeled $NY\Delta$) matter at
  temperature $T=0.1$~MeV. The remaining panels show the same
  dependence in neutrino trapped regime at temperature $T=50$~MeV in
  the two cases $Y_{L,\mu} = Y_{L,e} = 0.1$ (middle panel);
  $Y_{L,e}=0.1$, $Y_{L,\mu}=0.1$ (lower panel).  The case
  $Y_{L,e} = 0.1$ is characteristic of a BNS merger remnant, whereas the
  case $Y_{L,e} = 0.4$ for SN. }
\label{fig:EoS_T_const} 
\end{center}
\end{figure}

\begin{figure}[t] 
\begin{center}
\includegraphics[width=1.1\linewidth,keepaspectratio]{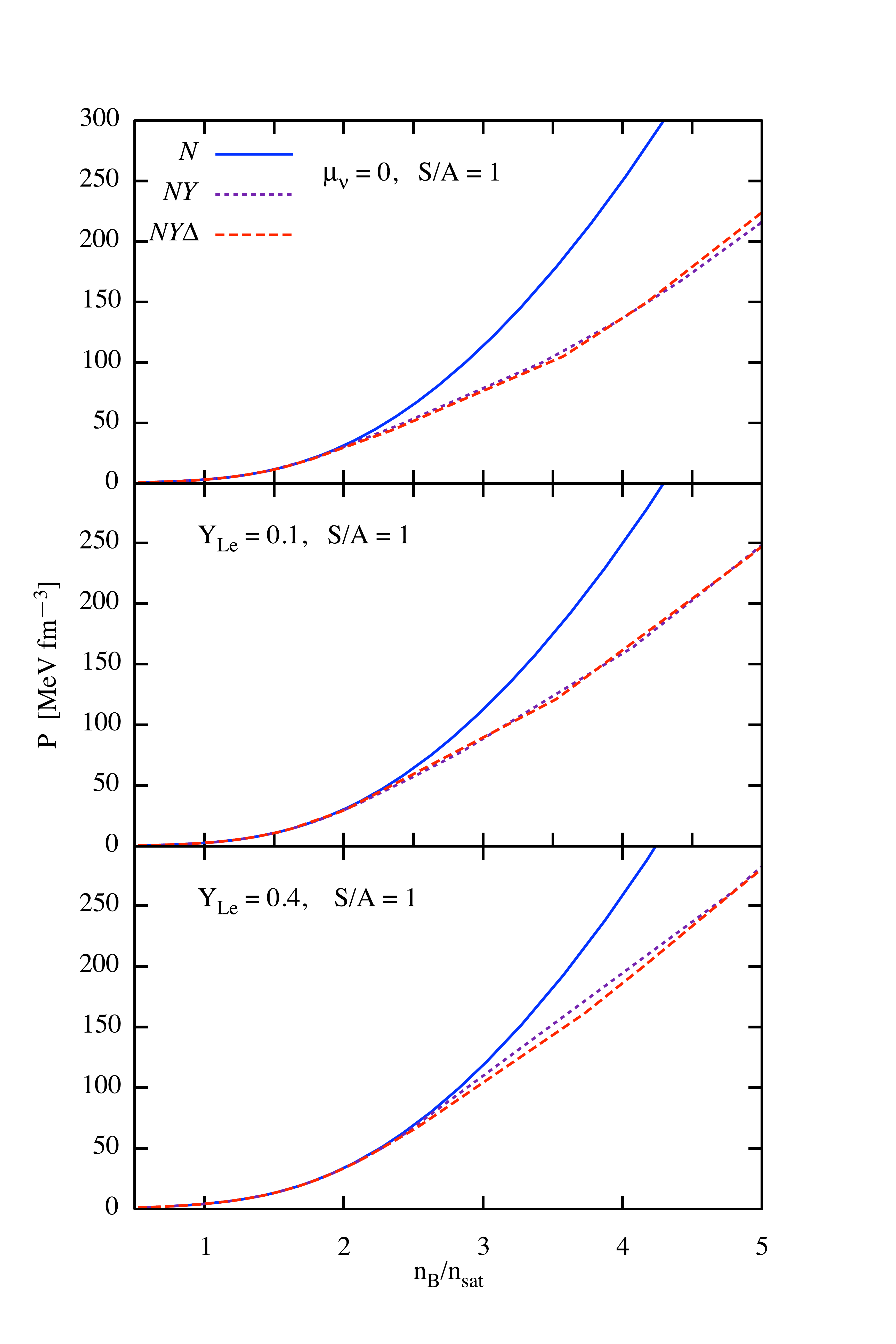}
\caption{Same as in Fig.~\ref{fig:EoS_T_const}, but in each panel
  instead of constant temperature the entropy per baryon
  is fixed at $S/A = 1$.
}
\label{fig:EoS_S_const} 
\end{center}
\end{figure}

We start the discussion with the finite-temperature EoS of dense
matter which is shown in Fig.~\ref{fig:EoS_T_const} for purely
nucleonic, hyperonic, and $\Delta$-admixed hyperonic matter.  It
includes the cases of low-temperature ($T=0.1$~MeV) matter in
$\beta$-equilibrium, which corresponds to the neutrino-free case.  In
addition it contains the results for $T=50$~MeV, which corresponds to
the trapped neutrino regime, for selected combinations of electron and
$\mu$-on fractions corresponding to SN physics ($Y_{L,\mu} = 0$,
$Y_{L,e} = 0.4$) and BNS merger physics ($Y_{L,\mu} =Y_{L,e} =
0.1$). The presence of hyperons strongly softens the EoS of nucleonic
matter consistent with the fact that new degrees of freedom appear
and, therefore, the degeneracy pressure of neutrons is reduced. If
$\Delta$'s are added to hypernuclear matter, the EoS becomes stiffer
in the high-density range and softer at the intermediate-density, as
already observed in the zero-temperature
calculations~\cite{Drago_PRC_2014,Cai_PRC_2015,Zhu_PRC_2016,Kolomeitsev2017,Sahoo_PRC_2018,Lijj2018b,Lijj2019,Ribes_2019}.
The case of constant entropy per baryon $S/A$ is shown in
Fig.~\ref{fig:EoS_S_const}. The softening of the EoS with the onset of
hyperons and $\Delta$-resonances is seen also in this case, as it is a
robust consequence of the onset of additional degrees of freedom.  At
asymptotically high density the $\Delta$ admixture leads (as in the
isothermal case) to a harder EoS.
%---------------------------------
% ----------------------------------------------------------------
%---------------------------------
% ---------

%-----------------------------------------------------------
\begin{figure}[t] 
\begin{center}
\includegraphics[width=1.1\linewidth,keepaspectratio]{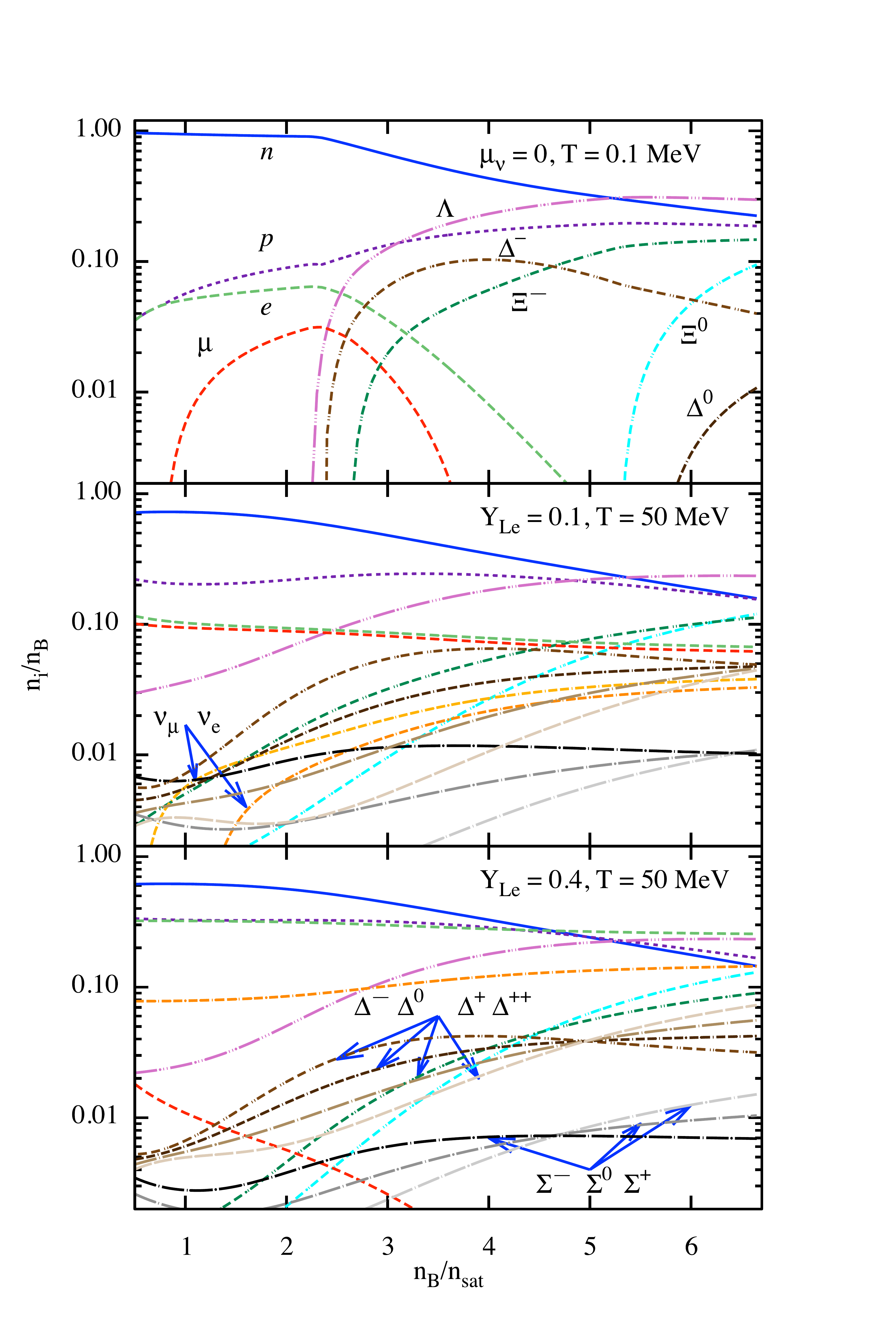}
\caption{ Composition of matter for $\Delta$-admixed hyperonic
  matter. The upper panel corresponds to $\beta$-equilibrium case at
  $T=0.1$~MeV whereas the remaining panels correspond to trapped
  neutrino matter at $T=50$~MeV with the lepton fractions fixed at
  $Y_{L,\mu} = Y_{L,e} = 0.1$ (middle panel), and $Y_{L,e}=0.4$,
  $Y_{L,\mu}=0$ (lower panel).  At $T=0.1$~MeV the hyperons
  $\Lambda$, $\Xi^-$ and $\Xi^0$ as well as $\Delta^-$ an $\Delta^0$
  resonances appear along with the
  standard nucleonic  (i.e. neutron-proton-electron and muon) composition.
  In the neutrino trapped regime the triplet $\Sigma^{0\pm}$ appears
  as well as electron and muon neutrinos (middle panel) and electron
  neutrinos (lower panel). }
\label{fig:abundances-T_const} 
\end{center}
\end{figure}
\begin{figure}[t] 
\begin{center}
\includegraphics[width=1.1\linewidth,keepaspectratio]{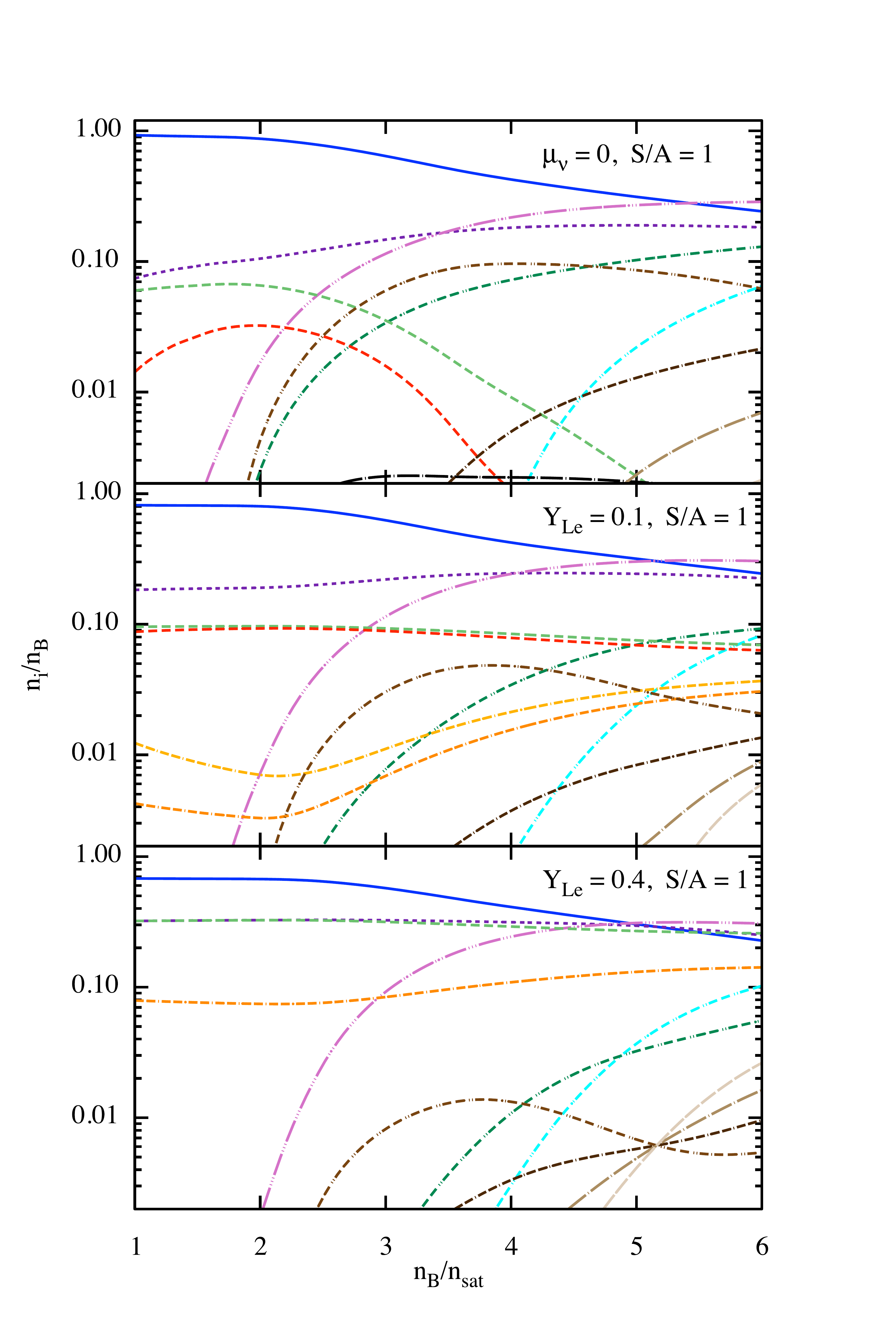}
\caption{ Same as in Fig.~\ref{fig:abundances-T_const}, but for
  constant entropy per baryon  $S/A = 1$.}
\label{fig:abundances-S_const} 
\end{center}
\end{figure}
%-----------------------------------------------------------

Figure \ref{fig:abundances-T_const} shows the composition of
finite-temperature hyperonic matter admixed with $\Delta$-resonances. 
Hyperons $\Lambda$, $\Xi^-$ and $\Xi^0$ appear in the given order 
at low temperature, with the $\Sigma^-$ hyperon fraction being 
strongly suppressed by the highly repulsive potential in nuclear matter at saturation
density~\cite{BartPhysRevLett,DOVER1984171,Maslov:2015wba,LopesPhysRevC2014,Gomes:2014aka,Miyatsu:2015kwa}.
This is in contrast with the early predictions made for the free
hyperonic gas, where $\Sigma^-$ was the first hyperon to
nucleate~\cite{Ambartsumyan1960SvA} and in an early version of the
present CDF work which employed a weaker repulsive
potential~\cite{Colucci2013}.  At finite temperature $T=50$~MeV the
isospin triplet of $\Sigma^{\pm,0}$ appears in amounts comparable (but
sub-leading) to other hyperons for both SN and BNS merger cases.

We find that among the $\Delta$-resonances only $\Delta^-$ and 
$\Delta^0$ appear in the matter (in the given order). At low
temperatures, there are clearly visible thresholds of appearance of
the resonances above twice the saturation density with only $\Delta^-$
reaching a significant (10\%) level in the intermediate density regime.
At high temperatures, the fractions of both $\Delta^-$ and $\Delta^0$
are comparable and phenomenologically significant $(\le 10\%)$.
The abundances of heavy baryons in the case of fixed entropy
 per baryon is shown in Fig.~\ref{fig:abundances-S_const}. Fixing the
 entropy requires the temperature to increase as the density increases
 and the abundances at each density correspond to a temperature that is 
 intermediate compared to those shown in Fig.~\ref{fig:abundances-T_const}. 
 As a consequence, for example, the thresholds for heavy baryons are 
 less steep and these are shifted to lower densities compared to the 
 $T=0.1$ MeV $\beta$-equilibrium case.

Let us turn now to the fractions of leptons. In the case of BNS
mergers the imposed condition $Y_{L,e} = Y_{L,\mu} = 0.1$ implies that
the fractions of electron and $\mu$-on are almost equal. The same 
applies to their neutrinos. In the SN case $Y_{L,\mu} = 0$, and the
$\mu$-on neutrinos are replaced by a much smaller amount of $\mu$-on
antineutrinos, which in turn allow for a small fraction of $\mu$-ons
to be present despite the condition $Y_{L,\mu} = 0$ was imposed. The
lepton fraction, which is intimately related to the charge neutrality 
condition, is affected once hyperons and $\Delta$-resonances are
introduced. The effect of adding $\Xi^-$ and $\Delta^-$ to the
composition in the low-temperature and $\beta$-equilibrated matter is
that the proton fraction becomes balanced by these particles rather
than leptons, and as a consequence the electron and $\mu$-on 
populations drop rapidly and eventually they become extinct at high 
densities. The decrease of lepton number densities with increasing
baryon density is observed also in the hot, $\beta$-equilibrated neutrino-trapped
matter, with the main difference being the fact that the lepton populations
remain finite at all densities. Since the lepton fractions are fixed in this case, 
this has the consequence that neutrino fractions increase with density. 
The effect of $\Xi^-$ and $\Delta^-$ at finite temperature is less
dramatic, since electrons are present at all densities, whereas the
fractions of $\mu$-ons depend on whether we adopt the BNS merger or SN
values of lepton fractions. In the first case, the electron and
$\mu$-on fractions are quantitatively close to each other. In the SN
case, $\mu$-on fraction is strongly reduced but is not zero because of
a population of $\mu$-on antineutrinos. 

 As well-known (see, e.g., Refs.~\cite{Lijj2019,Raduta2021PhLB}), the onset of $\Delta^-$ shifts
the balance between the chemical potentials of particles participating
in the Urca reactions $n\to p + e+ \bar \nu$ and $e+p \to n+ \nu$. The
proton fraction becomes large enough (compared to the $npe\mu$-matter)
so that the first Urca process can take place in the matter. The electron
extinction implies that the first Urca process is favored (because of
the absence of final state Pauli blocking of electron states) compared
to the second one, which is suppressed because of the absence of the
initial state electrons. Thus, the presence of $\Delta$'s can
  promote the nucleonic Urca processes. In addition, $\Delta$'s 
  themselves participate in Urca processes as their emergence can lead
  to additional processes, such as $\Delta^- \to n + e^- + \bar \nu$
  or $\Delta^- \to \Lambda + e^- + \bar \nu$~\cite{Prakash1992ApJ}. In
  hot matter these processes may contribute to the neutrino opacity,
  which is relevant in supernova
  context~\cite{Sumiyoshi2007,Fischer2009,OConnor_2011,Schneider2020},
  and bulk viscous damping of density oscillations in BNS
  mergers~\cite{Alford2019a,Alford2019b,Alford2020a,Alford2021a,Alford2021b}.

%---------------------------------
\begin{figure}[t] 
\begin{center}
\includegraphics[width=1.1\linewidth,keepaspectratio]{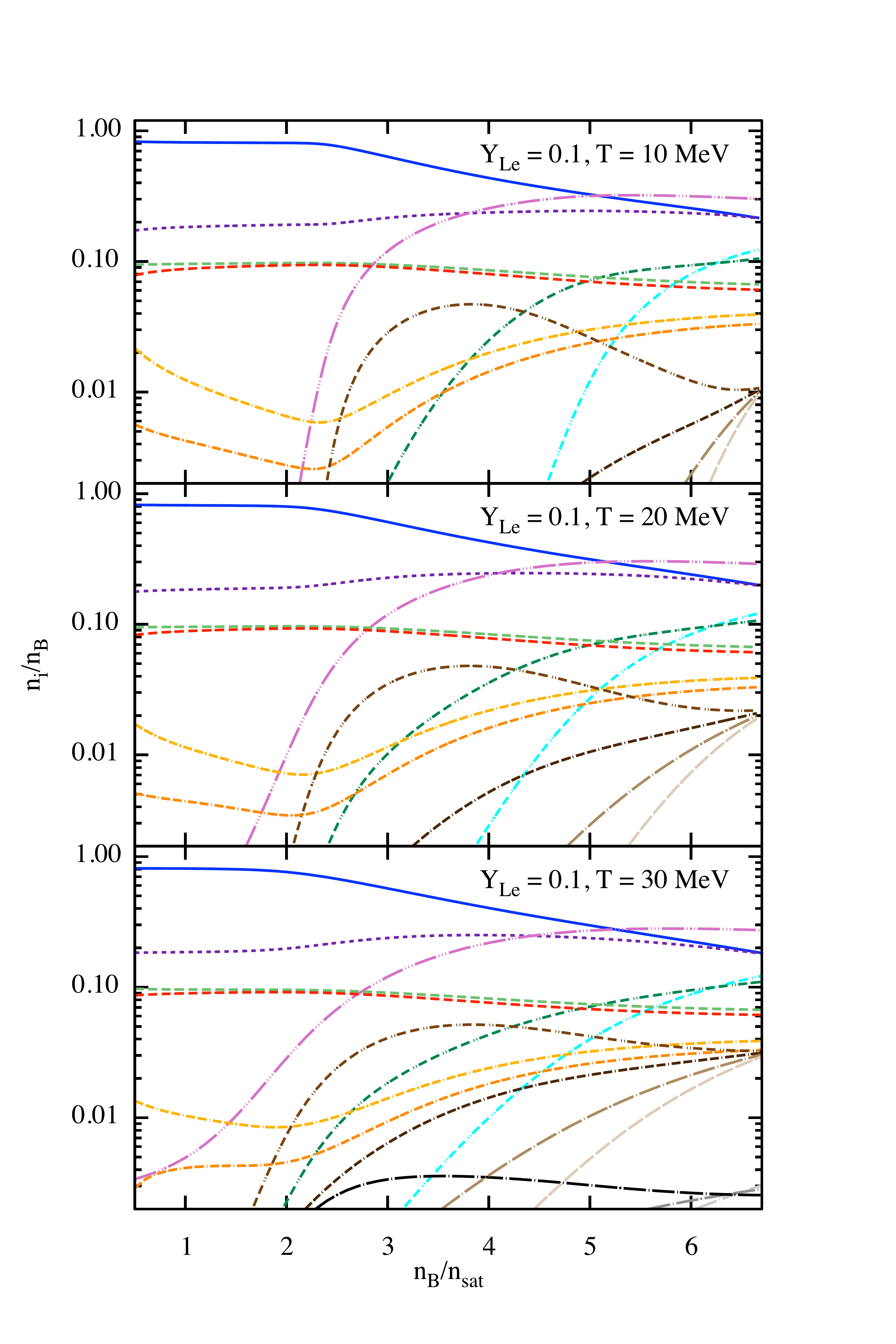}
\caption{Composition of $\Delta$-admixed hyperonic matter as in
  Fig.~\ref{fig:abundances-T_const}. The conventions are the same
  except that the lepton number fractions are fixed at $Y_{L,e}= Y_{L,\mu}=0.1$, 
  and the panels have fixed temperatures $T=10$, 20, and 30~MeV as 
  labeled. The lepton number fractions are characteristic of BNS mergers. 
}
\label{fig:abundances-hyperons-T_dep_Y_01} 
\end{center}
\end{figure}
% ---------------------------------
\begin{figure}[t] 
\begin{center}
\includegraphics[width=1.1\linewidth,keepaspectratio]{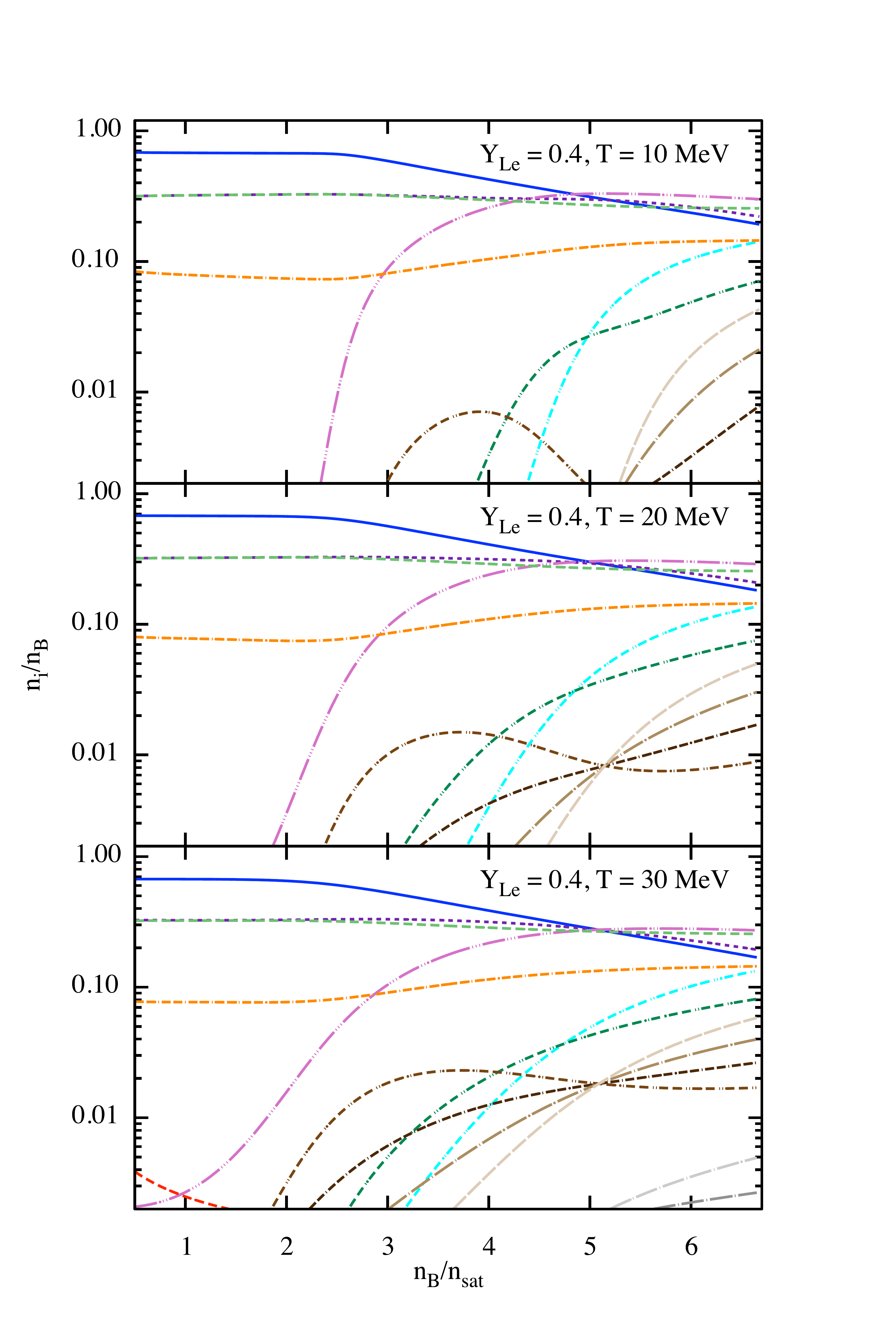}
\caption{Same as in Fig.~\ref{fig:abundances-hyperons-T_dep_Y_01}, but
  for lepton number fractions fixed at $Y_{L,e}= 0.4$ and $Y_{L,\mu}=0$ 
  which are characteristic of supernova matter. }
\label{fig:abundances-hyperons-T_dep_Y_04} 
\end{center}
\end{figure}
%---------------------------------

Let us finally comment on the high-temperature and low-density limit,
where our computations are limited to the density $0.5n_{\rm sat}$. 
It is seen from Fig. \ref{fig:abundances-T_const} that in this limit
hyperon and $\Delta$-resonance thresholds are absent and they
propagate up to the lower bound of the density range considered. A
hint on the presence of these species at lower densities is provided
by the recent observation that the low-density hot nuclear matter
contains a significant fraction of strangeness ($\Lambda$-particles)
as well as $\Delta$-resonances in addition to clusters and free
nucleons~\cite{Sedrakian:2020cjt}.

In closing our discussion of the numerical results, we would like to
explore the evolution of the fractions of particles with temperature
while keeping the lepton fractions constant according to
Eqs.~\eqref{eq:Y_BNS} and ~\eqref{eq:Y_SN}. Figures
\ref{fig:abundances-hyperons-T_dep_Y_01} and
\ref{fig:abundances-hyperons-T_dep_Y_04} below address these cases.

Consider first the case of BNS merger with fixed values of lepton
fractions $Y_{L,e}= Y_{L,\mu}=0.1$ and temperature values
$T= 10, 20$ and 30 MeV as shown in the
panels of Fig.~\ref{fig:abundances-hyperons-T_dep_Y_01}.
First note that the $n,p,e,\mu$ fractions depend weakly on
the temperature. The equality of lepton numbers implies that the
electron and $\mu$-on fractions are almost equal. The small electron
excess over $\mu$-ons is compensated by the $\mu$-on neutrino excess
over the electron-neutrinos. At high densities, the neutrino fractions 
depend weakly on the temperature as well but at low densities their
fractions decrease with increasing temperature and, eventually, become
negative at temperatures between 40 and 50 MeV.

%---------------------------------
\begin{figure}[t] 
\begin{center}
\includegraphics[width=1.1\linewidth,keepaspectratio]{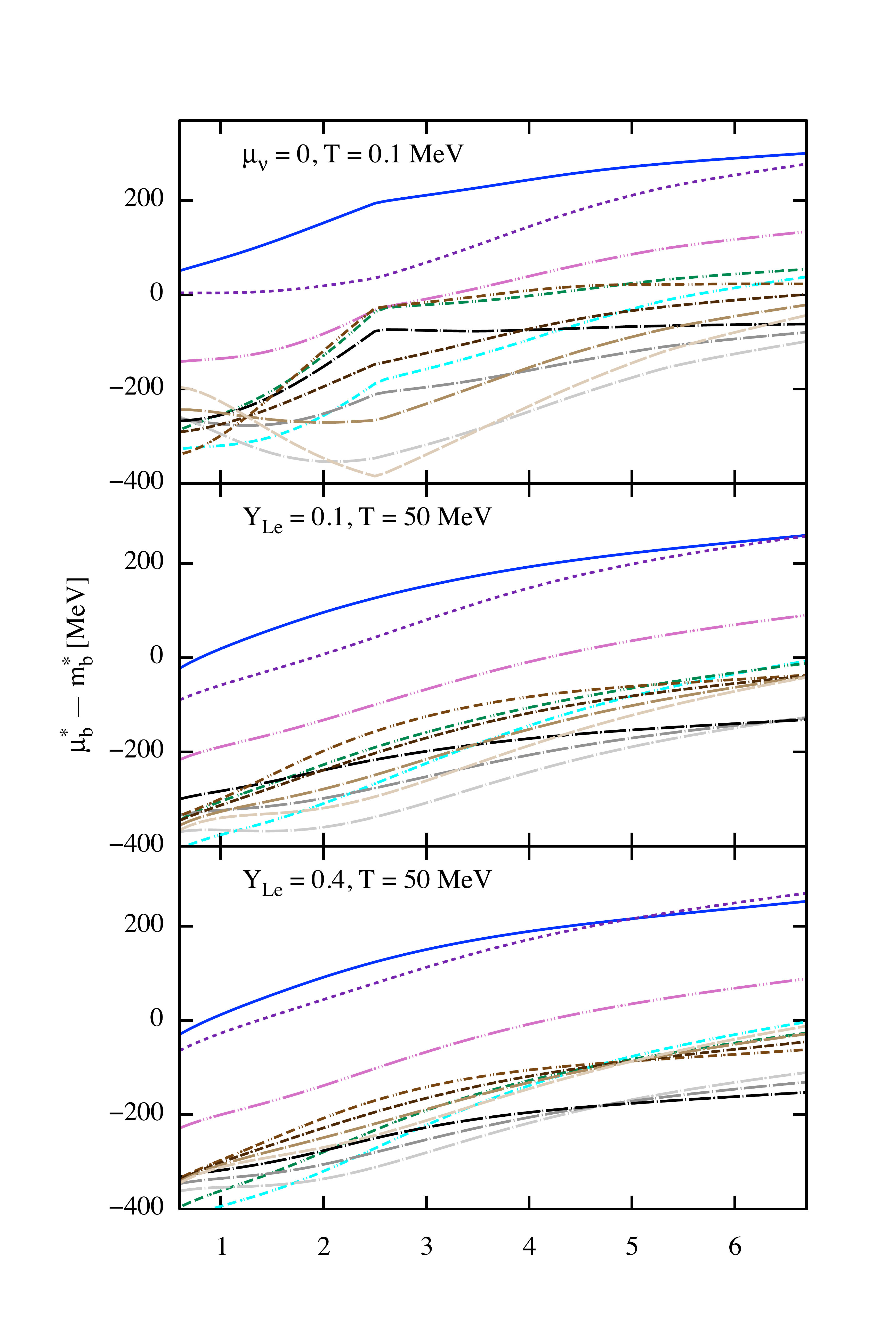}
\caption{Effective chemical potentials with the effective mass subtracted
  as functions of the normalized baryon density $n_B/n_{\rm sat}$.
  Each particles species are shown by the same lines as in
  Fig.~\ref{fig:abundances-T_const} with temperature and lepton
  fraction fixed as indicated. }
\label{fig:chempot_T_const} 
\end{center}
\end{figure}
% ---------------------------------
\begin{figure}[t] 
\begin{center}
\includegraphics[width=1.1\linewidth,keepaspectratio]{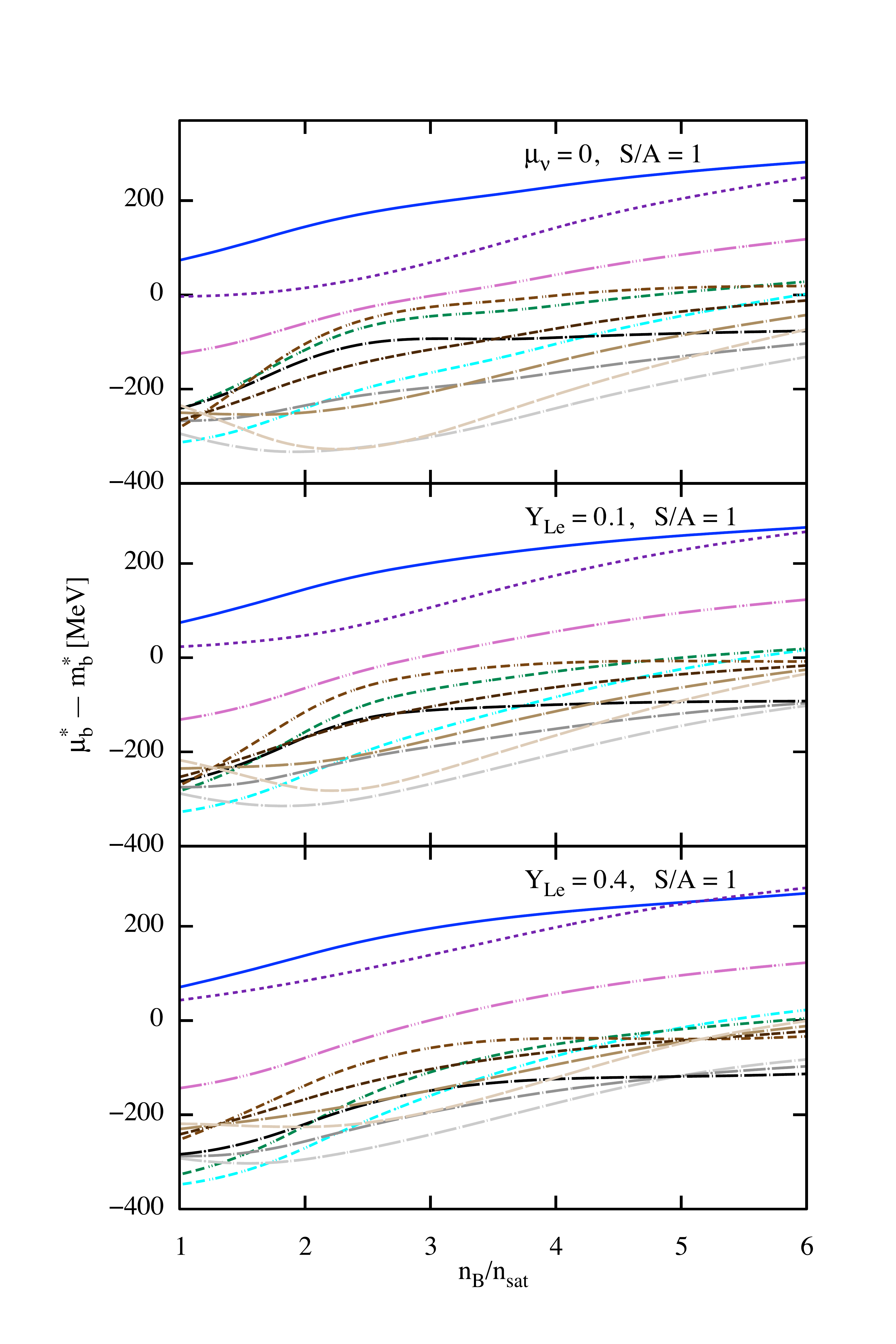}
\caption{Same as in Fig.~\ref{fig:chempot_T_const}, but for constant
  entropy per baryon $S/A = 1$. }
\label{fig:chempot_S_const} 
\end{center}
\end{figure}
%---------------------------------
Hyperons and $\Delta$-resonances have still sharply increasing
fractions at the thresholds at $T=10$~MeV similar to the
low-temperature, neutrino-free regime. The main change visible 
in the low-density regime is the shift of the thresholds of hyperons and
$\Delta$-resonances to lower densities with increasing temperature.
In the high-density regime, the changes with the temperature are 
not significant as the quantum degeneracy dominates thermal effects. 
In this density regime, the $\Lambda$ is the most abundant among the
heavy baryon species and its fraction exceeds that of neutron for
$n_B/n_{\rm sat} \ge 6.$ The $\Xi$-hyperons have similar to
  $\Delta^-$ fractions $\lesssim 10\%$, whereas the abundance of
  $\Delta^0$ is mildly suppressed at lower temperatures and becomes
  comparable to that of $\Delta^-$ at $T\geq 30$~MeV. Finally, for
$T\geq 20$~MeV the $\Sigma$ hyperons appear, but their fractions
remain below $1\%$.

Consider next the case of SN with fixed values of lepton fractions
$Y_{L,e}= 0.4$ and $Y_{L,\mu}=0$ as shown in
Fig.~\ref{fig:abundances-hyperons-T_dep_Y_04}. The remarks regarding
the shift of the heavy-baryon thresholds toward low densities with
increasing temperature are valid also in this case. At high densities
the dominance of $\Xi^0$ over $\Xi^-$ occurs
earlier, the reason being the suppression of $\Xi^-$ fractions by
electrons with preassigned lepton fraction. Indeed, electrons supply
the necessary negative charge which was otherwise due to $\Xi^-$
hyperons.  The main difference to the BNS case arises from the
suppression of the $\mu$-on fraction to below 1\%. Their non-vanishing 
number is due to the presence of $\mu$-on antineutrinos, as pointed
out above. Because of this, the charge neutrality is mainly maintained
by the equality of the abundances of protons and electrons, with
slight departure due to the presence of $\Xi^-$ and $\Delta^-$ at high
density. The small $\mu$-on fraction also results in the complete
dominance of the electron-neutrinos over their $\mu$-onic
counterparts.

As pointed out in the previous work~\cite{Sedrakian2021Univ}, where
resonances were neglected, there is a special {\it isospin degeneracy
  (ID)} point where the fractions within each isospin multiplet
coincide. We find that in the $\Delta$-resonance admixed matter this
feature is maintained and extended to the $\Delta^{\pm}\Delta^{++}$,
and $\Delta^0$ resonances.  At the ID point of the isospin multiplet
of $\Sigma^{0,\pm}$ hyperons the fractions of $\Sigma^-$ and
$\Sigma^+$ interchange their roles from being most abundant to least
abundant $\Sigma$-hyperon with increasing density. We see that the
fractions of $\Delta^{-,0}$ resonances, the fractions of isospin
multiplet of $\Sigma^{0,\pm}$ hyperons, $n$ and $p$ fractions, as well
as $\Xi^-$ and $\Xi^0$ fractions coincide at that point. This property
becomes evident from the $\beta$-equilibrium conditions
\eqref{eq:c1}--\eqref{eq:c4}. First, note that at high densities
$\mu_n^*-\mu_p^*\simeq \mu_n-\mu_p$ because the density
scaling~\eqref{eq:h_function_rho} implies that the contribution of the
$\rho$-meson mean-field to the effective baryon chemical
potentials~\eqref{eq:mu_eff} and \eqref{eq:mu_eff_d} vanishes
exponentially.  Now, if at any ID point the neutron and proton
fractions are equal, i.e., $\mu_n^*=\mu_p^*$, then the charge chemical
potential $\mu_Q=\mu_p-\mu_n = 0$ and, therefore, $\mu_b=\mu_B$. In
this case, the effective chemical potentials within any given
isospin-multiplet are equal and, therefore, their fractions are equal
as well.

% ---------------------------------
\begin{figure}[t]
\begin{center}
\includegraphics[width=\linewidth,keepaspectratio]{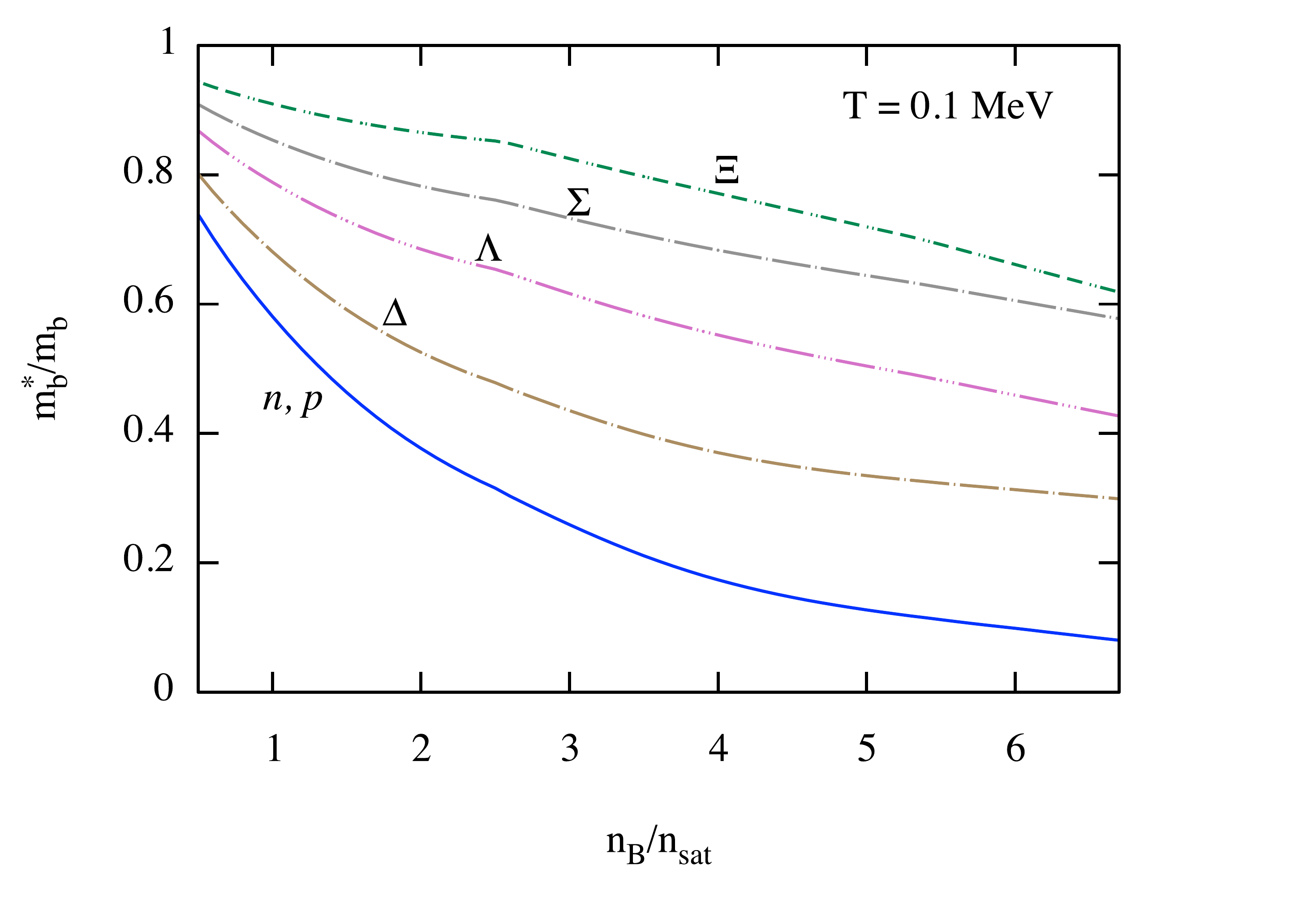}
\caption{Effective Dirac masses of baryons as functions of normalized by
  saturation density  baryonic density. The temperature is fixed at
  $T=0.1$~MeV for $\beta$-equilibrated, neutrino-free matter.
  Note that the isospin multiplets have 
  the same effective mass in the present model of CDF.
}
\label{fig:meff-hyperons}
\end{center}
\end{figure}
% -------------------------------------

Returning to the chemical potentials, we show in
Figs.~\ref{fig:chempot_T_const} and \ref{fig:chempot_S_const} their
effective values minus the respective effective masses for
$\Delta$-admixed hypernuclear matter in the cases of constant
temperature and fixed entropy per baryon, respectively. The emergence
of the isospin degeneracy point is seen in finite-temperature
neutrino-trapped matter calculations in each isospin multiplet for all
lepton number combinations considered. To the left from this point
$\mu_Q\leq 0$ which according to Eqs.~\eqref{eq:c1}--\eqref{eq:c4}
implies that baryons with smaller charges are more abundant. To the
right of this point $\mu_Q\geq 0$, and the ordering of baryon
fractions within each multiplet is reversed, as seen in
Fig.~\ref{fig:chempot_T_const}. The same feature is observed also in
the case of constant entropy (rather than constant temperature) case,
see Refs.~\cite{Malfatti:2019tpg,Raduta:2020fdn}.

Consider next the effective masses of the baryons, which are shown in
Fig.~\ref{fig:meff-hyperons} for $\beta$-equilibrium matter at fixed
temperature $T=0.1$~MeV as functions of density. The temperature
dependence of the effective masses of baryons is very weak. The
effective masses of isospin multiplets ($n,p$), $\Sigma^{0,\pm}$,
$\Xi^{0,-}$ and $\Delta^{0,-}$ are degenerate, which have important
implications for degeneracies in chemical potentials.

% ---------------------------------
\begin{figure}[t]
\begin{center}
\includegraphics[width=\linewidth,keepaspectratio]{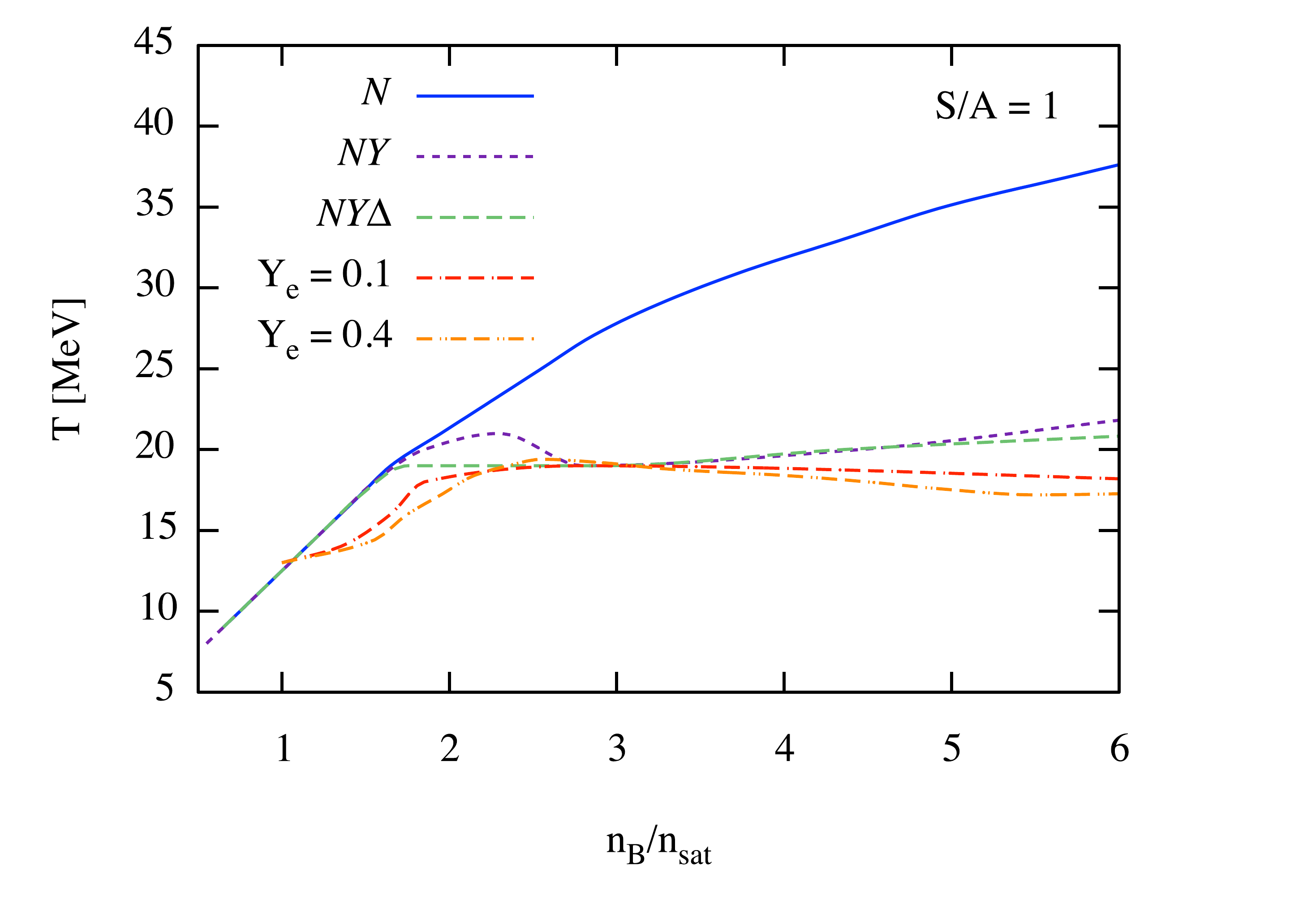}
\caption{ Dependence of temperature on density for fixed $S/A = 1$ in
  the cases of neutrino-transparent matter ($\mu_{\nu} =0$) with different
  compositions labeled as $N$, $NY$, and $NY\Delta$. The cases of
  neutrino-trapped matter with $Y_e=Y_{\mu} = 0.1$ and
  $Y_e=0.4, Y_{\mu}=0$ are shown shown by dash-dotted and
  dash-double-dotted curves.  }
\label{fig:T_rho}
\end{center}
\end{figure}
% -------------------------------------
The density dependence of the temperature for fixed $S/A=1$ for the
cases of $\mu_\nu=0$ and neutrino-trapped regime with
$Y_e=Y_{\mu} = 0.1$ and $Y_e=0.4, Y_{\mu}=0$ are shown in
Fig.~\ref{fig:T_rho}. The neutrino trapping occurs for temperatures
(roughly) $T\geq 5$~MeV~\cite{Alford2019a,Alford2019b,Alford2020a}. It
is seen that the temperature quickly rises above this limit as the
density increases. This implies that most of the volume of an
isentropic star with $S/A=1$ will be in the neutrino-trapped
regime. We, therefore, conclude that the upper panels in
Figs.~\ref{fig:abundances-S_const} and \ref{fig:chempot_T_const} do
not refer to a realistic situation to be encountered in BNS or SN
contexts.

\section{Cold and hot, isentropic compact stars}
\label{sec:MR}

Next, it is useful to use hot and cold EoS of $\Delta$-admixed 
hyperonic matter presented in the previous sections to compute the
spherical symmetrical static configurations of compact stars. The case
of cold EoS can be confronted with the current astrophysical constraints. 
Such analysis in the case of $\Delta$-admixed
hyperonic matter  can be found in Refs.~\cite{Lijj2018b,Lijj2019,
  Ribes_2019,Malfatti:2019tpg,Raduta:2020fdn}.  The astrophysical
constraints against which our EoS will be tested are as follows:
% -------------------------------------------------------------
%---------------------------------
\begin{figure*}[t] 
\begin{center}
\includegraphics[width=0.7\linewidth,keepaspectratio]{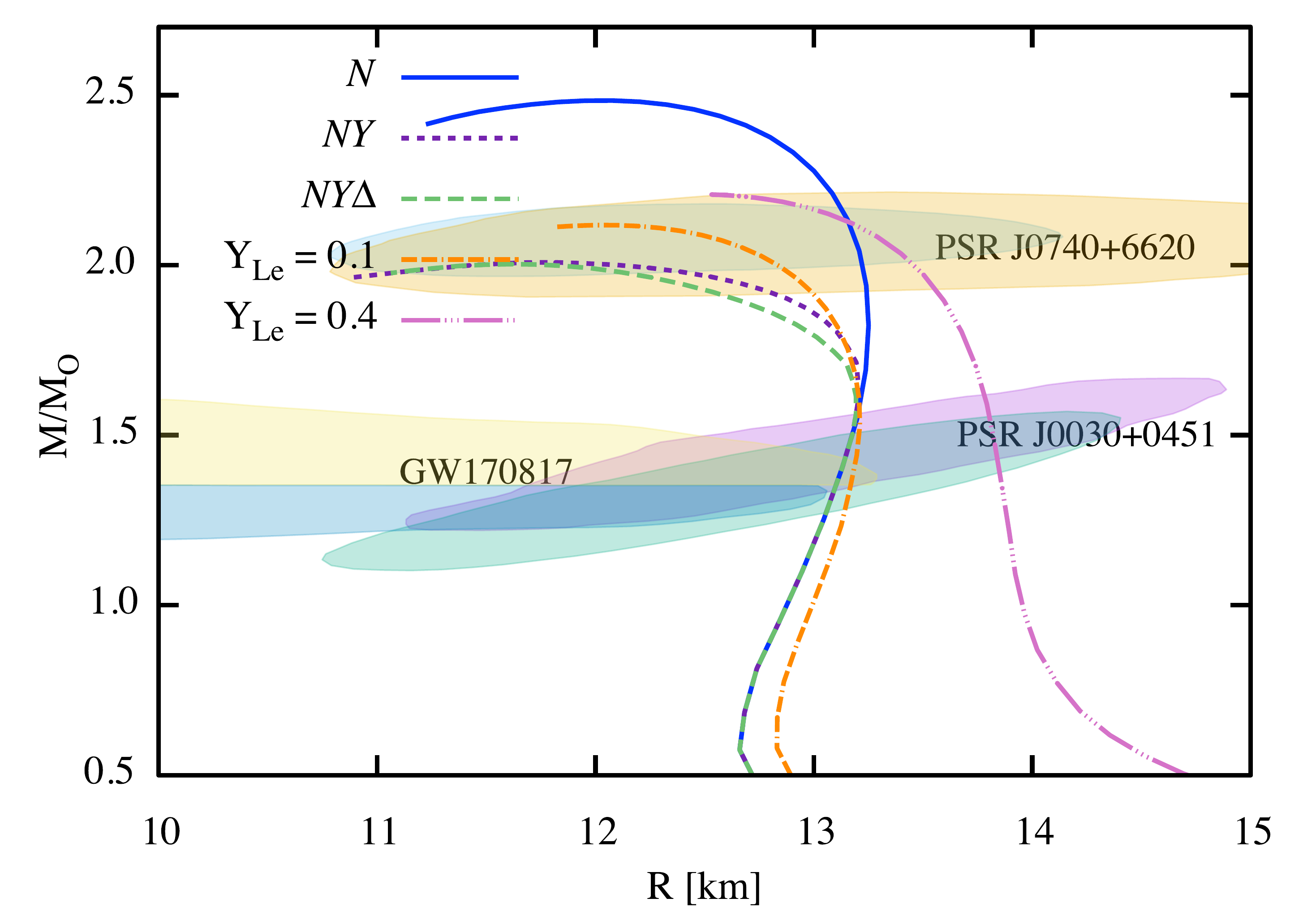}
\caption{Gravitational mass versus radius for non-rotating
  spherically-symmetric stars. Three sequences are shown for
  $\beta$-equilibrated, neutrino-transparent stars with nucleonic
  ($N$), hypernuclear ($NY$) and $\Delta$-admixed hypernuclear
  ($NY \Delta$) composition for $T=0.1$~MeV.  In addition, we show
  sequences of fixed $S/A=1$  neutrino-trapped, isentropic stars  
  composed of $NY \Delta$ matter in two cases of constant lepton
  fractions $Y_{Le}=Y_{L\mu}=0.1$ and $Y_{Le}=0.4,\,Y_{L\mu}=0$. The
  ellipses show 90\% CI regions for PSR J0030+0451, PSR
  J0740+6620 and gravitational wave event GW170817 (see the text for
  details).  }
\label{fig:MR} 
\end{center}
\end{figure*}
% -----------------------------------------------
\begin{description}
\item {\it (a) PSR J0030+0451} is the first object with highly
  accurate inferred mass and radius from X-ray
  observations~\cite{NICER2019a,NICER2019b}. Both the mass and the
  radius were inferred by fitting to the data obtained by the NICER
  X-ray observatory. The modeling of the soft $X$-ray pulses emitted by
  hot spots of a rotating star leads to two (independent) predictions
  ($68\%$ credible interval (CI)) $M=1.34^{+0.15}_{-0.16} M_{\odot}$,
  $R= 12.71^{+1.14}_{-1.19}$~km~\cite{NICER2019a} and
  $M = 1.44^{+0.15}_{-0.14}\, M_{\odot}$,
  $R = 13.02^{+1.24}_{^-1.06}$~km ~\cite{NICER2019b}.

\item {\it (b) PSR J0740+6620} is the second pulsar with a measured
  mass and radius again via observations and analysis of the NICER
  X-ray observatory data. Its radius estimates are
  $12.39^{+1.30}_{-0.98}$~\cite{NICER2021a} and
  $13.71^{+2.61}_{-1.50}$ km~\cite{NICER2021b} and the mass estimates
  are 
  $2.07^{+0.07}_{-0.07}\,M_{\odot}$ and
  $2.08^{+0.09}_{-0.09}$ $M_{\odot}$($68\%$ CI). Note that the mass of
  this pulsar was independently measured to be
  $2.08^{+0.07}_{-0.07}\,M_\odot$ using the effect of Shapiro
  delay~\citep{NANOGrav2019}.

\item {\it (c) GW170817} is the first multimessenger gravitational
  wave event which among various observables in the gravitational and
  electromagnetic spectrum allowed an inference of the tidal
  deformability of a star involved by the LIGO-Virgo
  Collaboration~\cite{LIGO_Virgo2019}.  The current upper limit of the
  (dimensionless) tidal deformability is $\tilde \Lambda \le 580$.
\end{description}
% ---------------------------------
\begin{table*}
  \begin{tabular}{ccccccccc}
    \hline
    & $M_{G,{\rm max}}$ &$R_{\rm max}$ & $n_{B,{\rm max}}/n_{\rm sat}$  
    
    & $n_{\Lambda}/n_{\rm sat}$ & $n_{\Xi^-}/n_{\rm sat}$
    & $n_{\Xi^0}/n_{\rm sat}$     & $n_{\Delta^-}/n_{\rm sat}$ & $n_{\Delta^0}/n_{\rm sat}$ \\
    &     $[M_{\odot}]$ &   [km]   &     & & &  \\\\
    %----------------------------------------------------------------------------------------------
    \hline
 $N$           & 2.48  & 12.1 &     7.27   & --& --&-- & --&  --\\
    %----------------------------------------------------------------------------------------------
 $NY$         & 2.01  & 11.8 &     7.79    &  2.25 & 2.55 &  5.62 & --  &  --\\
    %----------------------------------------------------------------------------------------------
 $NY\Delta$& 2.00  & 11.6 &     8.35    &  2.25  & 2.64 & 5.35 &  2.4 & 5.8 \\
    \hline
  \end{tabular}
  \caption{ Properties of non-rotating spherically symmetric cold
    $\beta$-equilibrated, neutrino-transparent, compact stars based on
    the EoS models considered in this work. The first three columns
    show the maximum gravitational mass ($M_{G,{\rm max}}$), and the
    corresponding radius $(R_{\rm max})$ and central baryon number density
    ($n_{c,{\rm max}}$) for the (cold) EoS with $N$, $NY$ and $NY\Delta$
    compositions.  The remaining columns show the threshold densities
    of heavy baryon defined here by the condition
    $n_{b,d}/n_{\rm sat}\ge 10^{-3}$.  The radius of a canonical mass
    $1.4M_{\odot}$ star for all cold EoS considered is
    $R_{1.4M_{\odot}} = 13.13$~km and its (dimensionless) tidal
    deformability is $\tilde \Lambda_{1.4M_{\odot}} = 707.3$.}
  \label{tab:NS}
  \end{table*}

  The 90\% CI ellipses for the constraints in the $M$-$R$ diagram are
  shown in Fig.~\ref{fig:MR}. The static solutions of Einstein's
  equations in spherical symmetry which are represented by the
  Tolman-Oppenheimer-Volkoff equations~\cite{Oppenheimer1939} were
  solved for input cold and finite-temperature isentropic EoS and the
  results are shown in Fig.~\ref{fig:MR}. In the case of cold EoS we
  consider the three cases of purely nucleonic ($N$), hyperonic ($YN$)
  and $\Delta$-resonance admixed hypernuclear matter ($YN\Delta$).
  Our nucleonic model has a maximum gravitational mass of
  $M_{\rm max}=2.48M_{\odot}$ and a radius of $R = 12.1$ km. For cold
  $YN$ and $YN\Delta$ matter the softening of the EoS results in a
  reduction of the maximum mass; for the $YN\Delta$ model
  $M_{\rm max}=2.0M_{\odot}$ with the corresponding value of the
  radius $R_{\rm max} = 11.6$ km. It is seen that the $M$-$R$ values
  of the models are compatible with the NICER inferences for canonical
  (i.e. $M\sim 1.4M_{\odot}$) and massive (i.e. $M\sim 2M_{\odot}$)
  compact stars. Note that the radii of all models are the same as in
  the case of a canonical mass star, as the onsets of the hyperons and
  $\Delta$-resonances are at densities that are beyond the central
  density of such a star. It is clearly seen that the bifurcation
  point of a heavy-baryon star from a purely nucleonic one lies at a
  higher mass $\sim 1.55M_{\odot}$. The results for the radius are
  compatible with the GW170817, but lie at the upper edge of the
  allowed radius. Also, the (dimensionless) tidal deformability of
  1.4$M_{\odot}$ star turns out to be
  $\tilde \Lambda_{1.4M_{\odot}} = 707.3$, which is larger than the
  upper limit $\tilde\Lambda \le 580$ given above.  However, note that
  larger values of $\Delta$-resonance-mesons couplings, which imply an
  early onset of these particles, can lead to a reduction of the
  radius of the star by about $15\%$ and a  reduction in tidal
  deformability; for a discussion of this point see
  Ref.~\cite{Lijj2019}. An alternative is a phase transition to
  quark matter phase at low densities~\cite{Li2020PhRvD}.

The case of hot, isentropic stars is relevant for transient states of
proto-neutron stars and BNS merger remnants. Fixing the value of the
entropy per particle and the lepton fraction at a constant value
throughout the entire star is clearly an approximation. These
quantities are known to have variations along with the radial profile of
the star. Nevertheless, such an assumption allows one to study (in a
first approximation) the effects of trapped neutrinos and temperature
on the configurations of stars. The sequences of isentropic stars
composed of $YN\Delta$ matter are shown in Fig.~\ref{fig:MR}. It is seen that
in the trapped neutrino regime the maximum masses of the stars are
shifted towards larger values.  At the same time, the radii of the
stars can be significantly larger than that of their cold
counterparts. For example, for a massive $M\sim 2M_{\odot}$ star this
difference is about 2 km for $Y_{Le}=0.4$.

\section{Conclusions}
\label{sec:conclusions}

We have extended our recent study of hypernuclear matter at finite
temperature in the neutrino-free and neutrino-trapped regimes to
include the non-strange $\Delta$-resonances. Our work is based on the
extension of the CDF formalism and numerical code  of
Ref.~\cite{Colucci2013} with the extensions in the hypernuclear sector described
previously in Ref.~\cite{Sedrakian2021Univ}. The zero-temperature counterpart
of this CDF and its astrophysical consequences were already discussed
extensively by Li et al.~\cite{Lijj2018b,Lijj2019,Li2020PhRvD}.  We
have exposed two physical cases by adjusting the lepton fractions of
electrons and $\mu$-ons to the conditions of BNS mergers and SN.

Firstly, we recovered the well-known features of the EoS that appear
when hyperons and $\Delta$-resonances are included in the composition
of matter.  The hyperonization softens the EoS compared to the
nucleonic case. Secondly, the inclusion of $\Delta$'s softens the EoS
at intermediate densities and stiffens it at high densities compared
to hypernuclear case (see Fig.~\ref{fig:EoS_T_const}).

Our conclusions can be summarized as follows:
\begin{itemize}

\item \underline{ $\Delta$-resonance thresholds.}  The
  zero-temperature abrupt increase in the heavy baryon
  abundances, in particular $\Delta$'s, at a given threshold is
  replaced by a much flatter increase at high temperatures
  with the low-density tail extending up to the lowest density value 
  considered $n = 0.5 n_{\rm sat}$, see Figs.~\ref{fig:abundances-T_const},
  \ref{fig:abundances-hyperons-T_dep_Y_01} and
  \ref{fig:abundances-hyperons-T_dep_Y_04}.  This clearly indicates
  that at finite temperatures the $\Delta$-resonances (in analogy to
  hyperons) extend further into the dilute gas regime of clustered
  nuclear matter.

\item \underline{Intermediate and large densities.}  The dominant
  $\Delta$-re\-so\-nance is $\Delta^-$ which has a threshold density
  close to that of $\Lambda$ for the moderate values of
  $\Delta$-couplings assumed in the present study. We find that
  $\Delta^{-}$ is the dominant charged heavy baryon at intermediate
  densities (up to $\sim 4n_{\rm sat}$) and becomes sub-dominant (but
  not significantly) to $\Xi^{-}$ at higher densities. The only other
  resonances are $\Delta^0$s, which appear at very high densities in
  cold, $\beta$-equilibrated matter, but their fractions are
  comparable to $\Delta^{-}$ if the matter is sufficiently hot, see
  Fig.~\ref{fig:abundances-T_const}.  The effect of the $\Delta$'s on
  the stiffening of the EoS of the hypernuclear matter at large
  densities, as seen in Figs. ~\ref{fig:EoS_T_const} and
  \ref{fig:EoS_S_const}, results in the increase of maximum masses of
  compact stars with $\Delta$'s compared to their hypernuclear
  counterparts, see Fig.~\ref{fig:MR}.

\item \underline{Neutrino species.} In the two considered
  astrophysical scenarios -- BNS mergers and SNs -- the neutrino
  populations differ considerably.  In the BNS merger scenario, the
  fractions of electron and $\mu$-on neutrinos are typically a few
  percent. The condition $Y_{L,e}= Y_{L,\mu} = 0.1$ enforces almost
  equal numbers of electrons and $\mu$-ons. In SN scenario the
  condition $Y_{L,\mu} = 0$ suppresses the $\mu$-ons leaving a
  negligible $\mu$-on fraction in the finite temperature
  neutrino-trapped regime due to a small amount of $\mu$-on
  antineutrinos. The matter is then dominated by electron neutrinos
  with a fraction $\sim 10\%$. The $\mu$-on antineutrino fraction is
  negligible.  The presence of $\Delta$ resonances will provide an
  additional source of interaction with neutrinos via direct Urca
  processes, e. g., $\Delta^-\to \Lambda +e^- +\bar\nu$. Consequently,
  the neutrino opacities may be affected by their very efficient
  direct-Urca coupling to $\Delta$'s (even in the absence of
  hyperons).  Furthermore, the bulk viscous damping of density
  oscillations in BNS mergers can be affected by the $\Delta$'s via
  non-equilibrium Urca processes involving $\Delta$'s.

\item \underline{Stellar configurations.} The stellar sequences based
  on the cold EoS of $NY\Delta$ matter are consistent with the
  astrophysical constraints set by the analysis of the NICER data on
  the masses and radii of PSR J0030+0451 and PSR J0740+6620. They are
  also consistent with the radius determination in the gravitational
  wave event GW170817, although the radii are at the upper edge of
  90\% CI region. Hot isentropic sequences can support larger masses
  than their cold counterparts. They are also more extended than the
  cold ones, the difference in the radii ranging up to a few km
  depending on the values of the entropy and lepton fraction and the
  mass range considered.
\end{itemize}

\section*{Acknowledgments}

We thank M. Alford, J.-J. Li, M. Oertel, A. Raduta and F. Weber for
discussion. 
This work was supported by the Volkswagen Foundation (Hannover,
Germany) grant No.  96 839, the Deutsche For\-schungs\-gemein\-schaft
(DFG) grant No.  SE1836/5-2, and the Polish National Science Centre
(NCN) grant 2020/\-37/B/ST9/01937.  The authors acknowledge the
support of the European COST Action “PHAROS”(CA16214).

%\bibliographystyle{JHEP}                                        
%\bibliography{Hyperons_ref,NYD}

\begin{thebibliography}{10}

\bibitem{Typel:2013rza}
S.~Typel, M.~Oertel and T.~Kl\"ahn, \emph{{CompOSE CompStar online supernova
  equations of state harmonising the concert of nuclear physics and
  astrophysics compose.obspm.fr}},
  \href{http://dx.doi.org/10.1134/S1063779615040061}{\emph{Phys. Part. Nucl.}
  {\bf 46} (2015) 633--664}, [\href{http://arxiv.org/abs/1307.5715}{{\tt
  1307.5715}}].

\bibitem{Prakash1997}
M.~{Prakash}, I.~{Bombaci}, M.~{Prakash}, P.~J. {Ellis}, J.~M. {Lattimer} and
  R.~{Knorren}, \emph{{Composition and structure of protoneutron stars}},
  \href{http://dx.doi.org/10.1016/S0370-1573(96)00023-3}{\emph{\physrep} {\bf
  280} (1997) 1--77}, [\href{http://arxiv.org/abs/nucl-th/9603042}{{\tt
  nucl-th/9603042}}].

\bibitem{Pons_ApJ_1999}
J.~A. Pons, S.~Reddy, M.~Prakash, J.~M. Lattimer and J.~A. Miralles,
  \emph{Evolution of proto{\textendash}neutron stars},
  \href{http://dx.doi.org/10.1086/306889}{\emph{ApJ} {\bf 513} (1999)
  780--804}.

\bibitem{Janka_PhysRep_2007}
H.~T. {Janka}, K.~{Langanke}, A.~{Marek}, G.~{Mart{\'\i}nez-Pinedo} and
  B.~{M{\"u}ller}, \emph{{Theory of core-collapse supernovae}},
  \href{http://dx.doi.org/10.1016/j.physrep.2007.02.002}{\emph{\physrep} {\bf
  442} (2007) 38--74}, [\href{http://arxiv.org/abs/astro-ph/0612072}{{\tt
  astro-ph/0612072}}].

\bibitem{Mezzacappa2015}
A.~{Mezzacappa}, E.~J. {Lentz}, S.~W. {Bruenn}, W.~R. {Hix}, O.~E.~B. {Messer},
  E.~{Endeve} et~al., \emph{{A Neutrino-Driven Core Collapse Supernova
  Explosion of a 15 M Star}}, {\emph{arXiv e-prints} (2015) arXiv:1507.05680},
  [\href{http://arxiv.org/abs/1507.05680}{{\tt 1507.05680}}].

\bibitem{Foglizzo2015}
T.~{Foglizzo}, R.~{Kazeroni}, J.~{Guilet}, F.~{Masset}, M.~{Gonz{\'a}lez},
  B.~K. {Krueger} et~al., \emph{{The Explosion Mechanism of Core-Collapse
  Supernovae: Progress in Supernova Theory and Experiments}},
  \href{http://dx.doi.org/10.1017/pasa.2015.9}{\emph{\pasa} {\bf 32} (2015)
  e009}, [\href{http://arxiv.org/abs/1501.01334}{{\tt 1501.01334}}].

\bibitem{Connor2018ApJ}
E.~P. {O'Connor} and S.~M. {Couch}, \emph{{Exploring Fundamentally
  Three-dimensional Phenomena in High-fidelity Simulations of Core-collapse
  Supernovae}}, \href{http://dx.doi.org/10.3847/1538-4357/aadcf7}{\emph{\apj}
  {\bf 865} (2018) 81}, [\href{http://arxiv.org/abs/1807.07579}{{\tt
  1807.07579}}].

\bibitem{Burrows2020MNRAS}
A.~{Burrows}, D.~{Radice}, D.~{Vartanyan}, H.~{Nagakura}, M.~A. {Skinner} and
  J.~C. {Dolence}, \emph{{The overarching framework of core-collapse supernova
  explosions as revealed by 3D FORNAX simulations}},
  \href{http://dx.doi.org/10.1093/mnras/stz3223}{\emph{\mnras} {\bf 491} (2020)
  2715--2735}, [\href{http://arxiv.org/abs/1909.04152}{{\tt 1909.04152}}].

\bibitem{Pascal2022}
A.~{Pascal}, J.~{Novak} and M.~{Oertel}, \emph{{Proto-neutron star evolution
  with improved charged-current neutrino-nucleon interactions}},
  \href{http://dx.doi.org/10.1093/mnras/stac016}{\emph{\mnras} {\bf 511} (2022)
  356--370}, [\href{http://arxiv.org/abs/2201.01955}{{\tt 2201.01955}}].

\bibitem{Sumiyoshi2007}
K.~{Sumiyoshi}, S.~{Yamada} and H.~{Suzuki}, \emph{{Dynamics and Neutrino
  Signal of Black Hole Formation in Nonrotating Failed Supernovae. I. Equation
  of State Dependence}}, \href{http://dx.doi.org/10.1086/520876}{\emph{\apj}
  {\bf 667} (2007) 382--394}, [\href{http://arxiv.org/abs/0706.3762}{{\tt
  0706.3762}}].

\bibitem{Fischer2009}
T.~{Fischer}, S.~C. {Whitehouse}, A.~{Mezzacappa}, F.~K. {Thielemann} and
  M.~{Liebend{\"o}rfer}, \emph{{The neutrino signal from protoneutron star
  accretion and black hole formation}},
  \href{http://dx.doi.org/10.1051/0004-6361/200811055}{\emph{\aap} {\bf 499}
  (2009) 1--15}, [\href{http://arxiv.org/abs/0809.5129}{{\tt 0809.5129}}].

\bibitem{OConnor_2011}
E.~O'Connor and C.~D. Ott, \emph{{Black Hole Formation in Failing Core-Collapse
  Supernovae}}, \href{http://dx.doi.org/10.1088/0004-637X/730/2/70}{\emph{\apj}
  {\bf 730} (2011) 70}, [\href{http://arxiv.org/abs/1010.5550}{{\tt
  1010.5550}}].

\bibitem{Schneider2020}
A.~{da Silva Schneider}, E.~{O'Connor}, E.~{Granqvist}, A.~{Betranhandy} and
  S.~M. {Couch}, \emph{{Equation of State and Progenitor Dependence of
  Stellar-mass Black Hole Formation}},
  \href{http://dx.doi.org/10.3847/1538-4357/ab8308}{\emph{\apj} {\bf 894}
  (2020) 4}, [\href{http://arxiv.org/abs/2001.10434}{{\tt 2001.10434}}].

\bibitem{Shibata_11}
K.~{Kyutoku}, M.~{Shibata} and K.~{Taniguchi}, \emph{{Coalescence of black
  hole-neutron star binaries}},
  \href{http://dx.doi.org/10.1007/s41114-021-00033-4}{\emph{Living Reviews in
  Relativity} {\bf 24} (2021) 5}, [\href{http://arxiv.org/abs/2110.06218}{{\tt
  2110.06218}}].

\bibitem{Rosswog2015}
S.~{Rosswog}, \emph{{The multi-messenger picture of compact binary mergers}},
  \href{http://dx.doi.org/10.1142/S0218271815300128}{\emph{International
  Journal of Modern Physics D} {\bf 24} (2015) 1530012--52},
  [\href{http://arxiv.org/abs/1501.02081}{{\tt 1501.02081}}].

\bibitem{Baiotti2017}
L.~{Baiotti} and L.~{Rezzolla}, \emph{{Binary neutron star mergers: a review of
  Einstein{\textquoteright}s richest laboratory}},
  \href{http://dx.doi.org/10.1088/1361-6633/aa67bb}{\emph{Reports on Progress
  in Physics} {\bf 80} (2017) 096901},
  [\href{http://arxiv.org/abs/1607.03540}{{\tt 1607.03540}}].

\bibitem{Baiotti:2019sew}
L.~{Baiotti}, \emph{{Gravitational waves from neutron star mergers and their
  relation to the nuclear equation of state}},
  \href{http://dx.doi.org/10.1016/j.ppnp.2019.103714}{\emph{Progress in
  Particle and Nuclear Physics} {\bf 109} (2019) 103714},
  [\href{http://arxiv.org/abs/1907.08534}{{\tt 1907.08534}}].

\bibitem{Hanau2019}
M.~Hanauske, J.~Steinheimer, A.~Motornenko, V.~Vovchenko, L.~Bovard, E.~R. Most
  et~al., \emph{Neutron star mergers: Probing the eos of hot, dense matter by
  gravitational waves},
  \href{http://dx.doi.org/10.3390/particles2010004}{\emph{Particles} {\bf 2}
  (2019) 44--56}.

\bibitem{Khadkikar:2021yrj}
S.~{Khadkikar}, A.~R. {Raduta}, M.~{Oertel} and A.~{Sedrakian}, \emph{{Maximum
  mass of compact stars from gravitational wave events with finite-temperature
  equations of state}},
  \href{http://dx.doi.org/10.1103/PhysRevC.103.055811}{\emph{\prc} {\bf 103}
  (2021) 055811}, [\href{http://arxiv.org/abs/2102.00988}{{\tt 2102.00988}}].

\bibitem{Alford2019a}
M.~G. {Alford} and S.~P. {Harris}, \emph{{Damping of density oscillations in
  neutrino-transparent nuclear matter}},
  \href{http://dx.doi.org/10.1103/PhysRevC.100.035803}{\emph{\prc} {\bf 100}
  (2019) 035803}, [\href{http://arxiv.org/abs/1907.03795}{{\tt 1907.03795}}].

\bibitem{Alford2019b}
M.~{Alford}, A.~{Harutyunyan} and A.~{Sedrakian}, \emph{{Bulk viscosity of
  baryonic matter with trapped neutrinos}},
  \href{http://dx.doi.org/10.1103/PhysRevD.100.103021}{\emph{\prd} {\bf 100}
  (2019) 103021}, [\href{http://arxiv.org/abs/1907.04192}{{\tt 1907.04192}}].

\bibitem{Alford2020a}
M.~{Alford}, A.~{Harutyunyan} and A.~{Sedrakian}, \emph{{Bulk Viscous Damping
  of Density Oscillations in Neutron Star Mergers}},
  \href{http://dx.doi.org/10.3390/particles3020034}{\emph{Particles} {\bf 3}
  (2020) 500--517}, [\href{http://arxiv.org/abs/2006.07975}{{\tt 2006.07975}}].

\bibitem{Alford2021a}
M.~{Alford}, A.~{Harutyunyan} and A.~{Sedrakian}, \emph{{Bulk viscosity from
  Urca processes: n p e {\ensuremath{\mu}} matter in the neutrino-trapped
  regime}}, \href{http://dx.doi.org/10.1103/PhysRevD.104.103027}{\emph{\prd}
  {\bf 104} (2021) 103027}, [\href{http://arxiv.org/abs/2108.07523}{{\tt
  2108.07523}}].

\bibitem{Alford2021b}
M.~G. {Alford} and A.~{Haber}, \emph{{Strangeness-changing rates and hyperonic
  bulk viscosity in neutron star mergers}},
  \href{http://dx.doi.org/10.1103/PhysRevC.103.045810}{\emph{\prc} {\bf 103}
  (2021) 045810}, [\href{http://arxiv.org/abs/2009.05181}{{\tt 2009.05181}}].

\bibitem{Alford2021c}
M.~G. {Alford}, A.~{Haber}, S.~P. {Harris} and Z.~{Zhang}, \emph{{Beta
  Equilibrium Under Neutron Star Merger Conditions}},
  \href{http://dx.doi.org/10.3390/universe7110399}{\emph{Universe} {\bf 7}
  (2021) 399}, [\href{http://arxiv.org/abs/2108.03324}{{\tt 2108.03324}}].

\bibitem{Most2022MNRAS}
E.~R. {Most}, S.~P. {Harris}, C.~{Plumberg}, M.~G. {Alford}, J.~{Noronha},
  J.~{Noronha-Hostler} et~al., \emph{{Projecting the likely importance of
  weak-interaction-driven bulk viscosity in neutron star mergers}},
  \href{http://dx.doi.org/10.1093/mnras/stab2793}{\emph{\mnras} {\bf 509}
  (2022) 1096--1108}, [\href{http://arxiv.org/abs/2107.05094}{{\tt
  2107.05094}}].

\bibitem{Celora:2022nbp}
T.~{Celora}, I.~{Hawke}, P.~C. {Hammond}, N.~{Andersson} and G.~L. {Comer},
  \emph{{Formulating bulk viscosity for neutron star simulations}},
  \href{http://dx.doi.org/10.1103/PhysRevD.105.103016}{\emph{\prd} {\bf 105}
  (2022) 103016}, [\href{http://arxiv.org/abs/2202.01576}{{\tt 2202.01576}}].

\bibitem{Alford2018b}
M.~G. {Alford} and S.~P. {Harris}, \emph{{{$\beta$} equilibrium in neutron-star
  mergers}}, \href{http://dx.doi.org/10.1103/PhysRevC.98.065806}{\emph{\prc}
  {\bf 98} (2018) 065806}, [\href{http://arxiv.org/abs/1803.00662}{{\tt
  1803.00662}}].

\bibitem{Sedrakian2021Univ}
A.~{Sedrakian} and A.~{Harutyunyan}, \emph{{Equation of State and Composition
  of Proto-Neutron Stars and Merger Remnants with Hyperons}},
  \href{http://dx.doi.org/10.3390/universe7100382}{\emph{Universe} {\bf 7}
  (2021) 382}, [\href{http://arxiv.org/abs/2109.01919}{{\tt 2109.01919}}].

\bibitem{Colucci2013}
G.~{Colucci} and A.~{Sedrakian}, \emph{{Equation of state of hypernuclear
  matter: Impact of hyperon-scalar-meson couplings}},
  \href{http://dx.doi.org/10.1103/PhysRevC.87.055806}{\emph{\prc} {\bf 87}
  (2013) 055806}, [\href{http://arxiv.org/abs/1302.6925}{{\tt 1302.6925}}].

\bibitem{Lijj2018b}
J.~J. {Li}, A.~{Sedrakian} and F.~{Weber}, \emph{{Competition between delta
  isobars and hyperons and properties of compact stars}},
  \href{http://dx.doi.org/10.1016/j.physletb.2018.06.051}{\emph{Physics Letters
  B} {\bf 783} (2018) 234--240}, [\href{http://arxiv.org/abs/1803.03661}{{\tt
  1803.03661}}].

\bibitem{Lijj2019}
J.~J. {Li} and A.~{Sedrakian}, \emph{{Implications from GW170817 for
  {\ensuremath{\Delta}}-isobar Admixed Hypernuclear Compact Stars}},
  \href{http://dx.doi.org/10.3847/2041-8213/ab1090}{\emph{\apjl} {\bf 874}
  (2019) L22}, [\href{http://arxiv.org/abs/1904.02006}{{\tt 1904.02006}}].

\bibitem{Li2020PhRvD}
J.~J. {Li}, A.~{Sedrakian} and M.~{Alford}, \emph{{Relativistic hybrid stars
  with sequential first-order phase transitions and heavy-baryon envelopes}},
  \href{http://dx.doi.org/10.1103/PhysRevD.101.063022}{\emph{\prd} {\bf 101}
  (2020) 063022}, [\href{http://arxiv.org/abs/1911.00276}{{\tt 1911.00276}}].

\bibitem{Lalazissis2005}
G.~A. {Lalazissis}, T.~{Nik{\v{s}}i{\'c}}, D.~{Vretenar} and P.~{Ring},
  \emph{{New relativistic mean-field interaction with density-dependent
  meson-nucleon couplings}},
  \href{http://dx.doi.org/10.1103/PhysRevC.71.024312}{\emph{\prc} {\bf 71}
  (2005) 024312}.

\bibitem{Schurhoff2010}
T.~{Sch{\"u}rhoff}, S.~{Schramm} and V.~{Dexheimer}, \emph{{Neutron Stars with
  Small Radii{\textemdash}The Role of {\ensuremath{\Delta}} Resonances}},
  \href{http://dx.doi.org/10.1088/2041-8205/724/1/L74}{\emph{\apjl} {\bf 724}
  (2010) L74--L77}, [\href{http://arxiv.org/abs/1008.0957}{{\tt 1008.0957}}].

\bibitem{Drago_PRC_2014}
A.~Drago, A.~Lavagno, G.~Pagliara and D.~Pigato, \emph{Early appearance of
  $\ensuremath{\Delta}$ isobars in neutron stars},
  \href{http://dx.doi.org/10.1103/PhysRevC.90.065809}{\emph{Phys. Rev. C} {\bf
  90} (2014) 065809}.

\bibitem{Cai_PRC_2015}
B.-J. Cai, F.~J. Fattoyev, B.-A. Li and W.~G. Newton, \emph{Critical density
  and impact of $\mathrm{\ensuremath{\Delta}}(1232)$ resonance formation in
  neutron stars},
  \href{http://dx.doi.org/10.1103/PhysRevC.92.015802}{\emph{Phys. Rev. C} {\bf
  92} (2015) 015802}, [\href{http://arxiv.org/abs/1501.01680}{{\tt
  1501.01680}}].

\bibitem{Zhu_PRC_2016}
Z.-Y. Zhu, A.~Li, J.-N. Hu and H.~Sagawa,
  \emph{$\mathrm{\ensuremath{\Delta}}(1232)$ effects in density-dependent
  relativistic hartree-fock theory and neutron stars},
  \href{http://dx.doi.org/10.1103/PhysRevC.94.045803}{\emph{Phys. Rev. C} {\bf
  94} (2016) 045803}.

\bibitem{Kolomeitsev2017}
E.~E. {Kolomeitsev}, K.~A. {Maslov} and D.~N. {Voskresensky}, \emph{{Delta
  isobars in relativistic mean-field models with {\ensuremath{\sigma}}-scaled
  hadron masses and couplings}},
  \href{http://dx.doi.org/10.1016/j.nuclphysa.2017.02.004}{\emph{\nphysa} {\bf
  961} (2017) 106--141}, [\href{http://arxiv.org/abs/1610.09746}{{\tt
  1610.09746}}].

\bibitem{Sahoo_PRC_2018}
H.~S. Sahoo, G.~Mitra, R.~Mishra, P.~K. Panda and B.-A. Li, \emph{Neutron star
  matter with $\mathrm{\ensuremath{\Delta}}$ isobars in a relativistic quark
  model}, \href{http://dx.doi.org/10.1103/PhysRevC.98.045801}{\emph{Phys. Rev.
  C} {\bf 98} (2018) 045801}.

\bibitem{Ribes_2019}
P.~Ribes, A.~Ramos, L.~Tolos, C.~Gonzalez-Boquera and M.~Centelles,
  \emph{{Interplay between $\Delta$ Particles and Hyperons in Neutron Stars}},
  \href{http://dx.doi.org/10.3847/1538-4357/ab3a93}{\emph{ApJ} {\bf 883} (2019)
  168}, [\href{http://arxiv.org/abs/1907.08583}{{\tt 1907.08583}}].

\bibitem{Sedrakian2021}
A.~{Sedrakian}, J.-J. {Li} and F.~{Weber}, \emph{{Hyperonization in compact
  stars}}, {\emph{arXiv e-prints} (2021) },
  [\href{http://arxiv.org/abs/2105.14050}{{\tt 2105.14050}}].

\bibitem{Malfatti:2019tpg}
G.~Malfatti, M.~G. Orsaria, G.~A. Contrera, F.~Weber and I.~F. Ranea-Sandoval,
  \emph{{Hot quark matter and (proto-) neutron stars}},
  \href{http://dx.doi.org/10.1103/PhysRevC.100.015803}{\emph{Phys. Rev. C} {\bf
  100} (2019) 015803}, [\href{http://arxiv.org/abs/1907.06597}{{\tt
  1907.06597}}].

\bibitem{Spinella2020:WSBook}
W.~M. Spinella and F.~Weber, \emph{Dense Baryonic Matter in the Cores of
  Neutron Stars, in: Topics on Strong Gravity}, pp.~85--152.
\newblock World Scientific, 2020.
\newblock 10.1142/11186.

\bibitem{Raduta:2020fdn}
A.~R. {Raduta}, M.~{Oertel} and A.~{Sedrakian}, \emph{{Proto-neutron stars with
  heavy baryons and universal relations}},
  \href{http://dx.doi.org/10.1093/mnras/staa2491}{\emph{\mnras} {\bf 499}
  (2020) 914--931}, [\href{http://arxiv.org/abs/2008.00213}{{\tt 2008.00213}}].

\bibitem{Sinha2013}
M.~{Sinha}, B.~{Mukhopadhyay} and A.~{Sedrakian}, \emph{{Hypernuclear matter in
  strong magnetic field}},
  \href{http://dx.doi.org/10.1016/j.nuclphysa.2012.12.076}{\emph{\nphysa} {\bf
  898} (2013) 43--58}, [\href{http://arxiv.org/abs/1005.4995}{{\tt
  1005.4995}}].

\bibitem{Thapa:2020ohp}
V.~B. {Thapa}, M.~{Sinha}, J.~J. {Li} and A.~{Sedrakian}, \emph{{Massive
  {\ensuremath{\Delta}} -resonance admixed hypernuclear stars with antikaon
  condensations}},
  \href{http://dx.doi.org/10.1103/PhysRevD.103.063004}{\emph{\prd} {\bf 103}
  (2021) 063004}.

\bibitem{Dexheimer:2021sxs}
V.~{Dexheimer}, K.~D. {Marquez} and D.~P. {Menezes}, \emph{{Delta baryons in
  neutron-star matter under strong magnetic fields}},
  \href{http://dx.doi.org/10.1140/epja/s10050-021-00532-6}{\emph{European
  Physical Journal A} {\bf 57} (2021) 216},
  [\href{http://arxiv.org/abs/2103.09855}{{\tt 2103.09855}}].

\bibitem{Typelparticles2018}
S.~Typel, \emph{Relativistic mean-field models with different parametrizations
  of density dependent couplings},
  \href{http://dx.doi.org/10.3390/particles1010002}{\emph{Particles} {\bf 1}
  (2018) 3--22}.

\bibitem{Swart1963}
J.~J. {de Swart}, \emph{{The Octet Model and its Clebsch-Gordan Coefficients}},
  \href{http://dx.doi.org/10.1103/RevModPhys.35.916}{\emph{Rev. Mod. Phys.}
  {\bf 35} (1963) 916--939}.

\bibitem{Dalen2014}
E.~N.~E. {van Dalen}, G.~{Colucci} and A.~{Sedrakian}, \emph{{Constraining
  hypernuclear density functional with {\ensuremath{\Lambda}}-hypernuclei and
  compact stars}},
  \href{http://dx.doi.org/10.1016/j.physletb.2014.06.002}{\emph{Phys. Lett. B}
  {\bf 734} (2014) 383--387}, [\href{http://arxiv.org/abs/1406.0744}{{\tt
  1406.0744}}].

\bibitem{Friedman2021PhLB}
E.~{Friedman} and A.~{Gal}, \emph{{Constraints on {\ensuremath{\Xi}}$^{-}$
  nuclear interactions from capture events in emulsion}},
  \href{http://dx.doi.org/10.1016/j.physletb.2021.136555}{\emph{Physics Letters
  B} {\bf 820} (2021) 136555}, [\href{http://arxiv.org/abs/2104.00421}{{\tt
  2104.00421}}].

\bibitem{Harada2021PhRvC}
T.~{Harada} and Y.~{Hirabayashi}, \emph{{{\ensuremath{\Xi}} -nucleus potential
  for {\ensuremath{\Xi}}$^{-}$ quasifree production in the
  $^{9}$Be(K$^{-}$,K$^{+}$) reaction}},
  \href{http://dx.doi.org/10.1103/PhysRevC.103.024605}{\emph{\prc} {\bf 103}
  (2021) 024605}, [\href{http://arxiv.org/abs/2101.00855}{{\tt 2101.00855}}].

\bibitem{Sasaki:2019qnh}
{\scshape HAL QCD} collaboration, K.~Sasaki et~al., \emph{{$\Lambda\Lambda$ and
  N$\Xi$ interactions from Lattice QCD near the physical point}},
  \href{http://dx.doi.org/10.1016/j.nuclphysa.2020.121737}{\emph{Nucl. Phys. A}
  {\bf 998} (2020) 121737}, [\href{http://arxiv.org/abs/1912.08630}{{\tt
  1912.08630}}].

\bibitem{Li2019PhysRevC}
J.~J. Li and A.~Sedrakian, \emph{Constraining compact star properties with
  nuclear saturation parameters}, {\emph{Phys. Rev. C} {\bf 100} (2019)
  015809}, [\href{http://arxiv.org/abs/1903.06057}{{\tt 1903.06057}}].

\bibitem{Raduta2021PhLB}
A.~R. {Raduta}, \emph{{{\ensuremath{\Delta}}-admixed neutron stars: Spinodal
  instabilities and dUrca processes}},
  \href{http://dx.doi.org/10.1016/j.physletb.2021.136070}{\emph{Physics Letters
  B} {\bf 814} (2021) 136070}, [\href{http://arxiv.org/abs/2101.03718}{{\tt
  2101.03718}}].

\bibitem{Bollig:2017lki}
R.~Bollig, H.~T. Janka, A.~Lohs, G.~Martinez-Pinedo, C.~J. Horowitz and
  T.~Melson, \emph{{Muon Creation in Supernova Matter Facilitates
  Neutrino-driven Explosions}},
  \href{http://dx.doi.org/10.1103/PhysRevLett.119.242702}{\emph{Phys. Rev.
  Lett.} {\bf 119} (2017) 242702}, [\href{http://arxiv.org/abs/1706.04630}{{\tt
  1706.04630}}].

\bibitem{Guo:2020tgx}
G.~Guo, G.~Mart\'\i{}nez-Pinedo, A.~Lohs and T.~Fischer, \emph{{Charged-Current
  Muonic Reactions in Core-Collapse Supernovae}},
  \href{http://dx.doi.org/10.1103/PhysRevD.102.023037}{\emph{Phys. Rev. D} {\bf
  102} (2020) 023037}, [\href{http://arxiv.org/abs/2006.12051}{{\tt
  2006.12051}}].

\bibitem{Perego2019EPJA}
A.~{Perego}, S.~{Bernuzzi} and D.~{Radice}, \emph{{Thermodynamics conditions of
  matter in neutron star mergers}},
  \href{http://dx.doi.org/10.1140/epja/i2019-12810-7}{\emph{European Physical
  Journal A} {\bf 55} (2019) 124}, [\href{http://arxiv.org/abs/1903.07898}{{\tt
  1903.07898}}].

\bibitem{Nakazato2022ApJ}
K.~{Nakazato}, F.~{Nakanishi}, M.~{Harada}, Y.~{Koshio}, Y.~{Suwa},
  K.~{Sumiyoshi} et~al., \emph{{Observing Supernova Neutrino Light Curves with
  Super-Kamiokande. II. Impact of the Nuclear Equation of State}},
  \href{http://dx.doi.org/10.3847/1538-4357/ac3ae2}{\emph{\apj} {\bf 925}
  (2022) 98}, [\href{http://arxiv.org/abs/2108.03009}{{\tt 2108.03009}}].

\bibitem{BartPhysRevLett}
S.~Bart, R.~E. Chrien, W.~A. Franklin, T.~Fukuda, R.~S. Hayano, K.~Hicks
  et~al., \emph{$\sigma$ hyperons in the nucleus},
  \href{http://dx.doi.org/10.1103/PhysRevLett.83.5238}{\emph{Phys. Rev. Lett.}
  {\bf 83} (1999) 5238--5241}.

\bibitem{DOVER1984171}
C.~Dover and A.~Gal, \emph{Hyperon-nucleus potentials},
  \href{http://dx.doi.org/https://doi.org/10.1016/0146-6410(84)90004-8}{\emph{Prog.
  Part. Nucl. Phys.} {\bf 12} (1984) 171--239}.

\bibitem{Maslov:2015wba}
K.~A. {Maslov}, E.~E. {Kolomeitsev} and D.~N. {Voskresensky},
  \emph{{Relativistic mean-field models with scaled hadron masses and
  couplings: Hyperons and maximum neutron star mass}},
  \href{http://dx.doi.org/10.1016/j.nuclphysa.2016.03.011}{\emph{\nphysa} {\bf
  950} (2016) 64--109}, [\href{http://arxiv.org/abs/1509.02538}{{\tt
  1509.02538}}].

\bibitem{LopesPhysRevC2014}
L.~L. Lopes and D.~P. Menezes, \emph{Hypernuclear matter in a complete su(3)
  symmetry group},
  \href{http://dx.doi.org/10.1103/PhysRevC.89.025805}{\emph{Phys. Rev. C} {\bf
  89} (2014) 025805}, [\href{http://arxiv.org/abs/1309.4173}{{\tt 1309.4173}}].

\bibitem{Gomes:2014aka}
R.~O. Gomes, V.~Dexheimer, S.~Schramm and C.~A.~Z. Vasconcellos,
  \emph{Many-body forces in the equation of state of hyperonic matter},
  \href{http://dx.doi.org/10.1088/0004-637x/808/1/8}{\emph{Ap. J.} {\bf 808}
  (2015) 8}.

\bibitem{Miyatsu:2015kwa}
T.~{Miyatsu}, M.-K. {Cheoun} and K.~{Saito}, \emph{{Equation of State for
  Neutron Stars with Hyperons and Quarks in the Relativistic Hartree-Fock
  Approximation}},
  \href{http://dx.doi.org/10.1088/0004-637X/813/2/135}{\emph{\apj} {\bf 813}
  (2015) 135}, [\href{http://arxiv.org/abs/1506.05552}{{\tt 1506.05552}}].

\bibitem{Ambartsumyan1960SvA}
V.~A. {Ambartsumyan} and G.~S. {Saakyan}, \emph{{The Degenerate Superdense Gas
  of Elementary Particles}}, {\emph{\sovast} {\bf 4} (1960) 187}.

\bibitem{Prakash1992ApJ}
M.~{Prakash}, M.~{Prakash}, J.~M. {Lattimer} and C.~J. {Pethick}, \emph{{Rapid
  Cooling of Neutron Stars by Hyperons and Delta Isobars}},
  \href{http://dx.doi.org/10.1086/186376}{\emph{\apjl} {\bf 390} (1992) L77}.

\bibitem{Sedrakian:2020cjt}
A.~{Sedrakian}, \emph{{Light clusters in dilute heavy-baryon admixed nuclear
  matter}},
  \href{http://dx.doi.org/10.1140/epja/s10050-020-00262-1}{\emph{European
  Physical Journal A} {\bf 56} (2020) 258},
  [\href{http://arxiv.org/abs/2009.00357}{{\tt 2009.00357}}].

\bibitem{NICER2019a}
T.~E. Riley, A.~L. Watts, S.~Bogdanov et~al., \emph{{A $NICER$ View of PSR
  J0030+0451: Millisecond Pulsar Parameter Estimation}},
  \href{http://dx.doi.org/10.3847/2041-8213/ab481c}{\emph{Astrophys. J. Lett.}
  {\bf 887} (2019) L21}, [\href{http://arxiv.org/abs/1912.05702}{{\tt
  1912.05702}}].

\bibitem{NICER2019b}
M.~C. Miller, F.~K. Lamb, A.~J. Dittmann et~al., \emph{{PSR J0030+0451 Mass and
  Radius from $NICER$ Data and Implications for the Properties of Neutron Star
  Matter}}, \href{http://dx.doi.org/10.3847/2041-8213/ab50c5}{\emph{Astrophys.
  J. Lett.} {\bf 887} (2019) L24}, [\href{http://arxiv.org/abs/1912.05705}{{\tt
  1912.05705}}].

\bibitem{NICER2021a}
T.~E. Riley, A.~L. Watts, P.~S. Ray et~al., \emph{{A NICER View of the Massive
  Pulsar PSR J0740+6620 Informed by Radio Timing and XMM-Newton Spectroscopy}},
  \href{http://dx.doi.org/10.3847/2041-8213/ac0a81}{\emph{Astrophys. J. Lett.}
  {\bf 918} (2021) L27}, [\href{http://arxiv.org/abs/2105.06980}{{\tt
  2105.06980}}].

\bibitem{NICER2021b}
M.~C. Miller, F.~K. Lamb, A.~J. Dittmann et~al., \emph{{The Radius of PSR
  J0740+6620 from NICER and XMM-Newton Data}},
  \href{http://dx.doi.org/10.3847/2041-8213/ac089b}{\emph{Astrophys. J. Lett.}
  {\bf 918} (2021) L28}, [\href{http://arxiv.org/abs/2105.06979}{{\tt
  2105.06979}}].

\bibitem{NANOGrav2019}
{\scshape NANOGrav} collaboration, H.~T. Cromartie, E.~Fonseca, S.~M. Ransom
  et~al., \emph{{Relativistic Shapiro delay measurements of an extremely
  massive millisecond pulsar}},
  \href{http://dx.doi.org/10.1038/s41550-019-0880-2}{\emph{Nat. Astron.} {\bf
  4} (2020) 72--76}, [\href{http://arxiv.org/abs/1904.06759}{{\tt
  1904.06759}}].

\bibitem{LIGO_Virgo2019}
{\scshape LIGO Scientific, Virgo} collaboration, B.~P. Abbott, R.~Abbott, T.~D.
  Abbot et~al., \emph{{Properties of the binary neutron star merger GW170817}},
  \href{http://dx.doi.org/10.1103/PhysRevX.9.011001}{\emph{Phys. Rev. X} {\bf
  9} (2019) 011001}, [\href{http://arxiv.org/abs/1805.11579}{{\tt
  1805.11579}}].

\bibitem{Oppenheimer1939}
J.~R. Oppenheimer and G.~M. Volkoff, \emph{On massive neutron cores},
  {\emph{Phys. Rev.} {\bf 55} (1939) 374--381}.

\end{thebibliography}

\providecommand{\href}[2]{#2}\begingroup\raggedright\endgroup

\end{document}